\documentclass[apj]{emulateapj}

\shorttitle{Inner Halo Streamers in the Era of Gaia}
\shortauthors{Re Fiorentin et al.}

\begin{document}

\title{New Signatures of the Milky Way Formation in the Local Halo and Inner Halo Streamers in the Era of Gaia}

\author{Paola Re Fiorentin$^1$}
\email{re$_$fiorentin@oato.inaf.it}
\author{Mario G. Lattanzi$^{1,2}$}
\author{Alessandro Spagna$^1$}
\author{Anna Curir$^1$}
\affil{$^1$INAF - Osservatorio Astrofisico di Torino, Strada Osservatorio 20, 10025 Pino Torinese, TO, Italy \\
$^2$Shanghai Astronomical Observatory, Chinese Academy of Sciences, 80 Nandan Road, 200030 Shanghai, China}

%\author{Paola Re Fiorentin}
%\affil{
%INAF - Osservatorio Astrofisico di Torino, Strada Osservatorio 20, 10025 Pino Torinese, TO, Italy}
%\email{re$_{-}$fiorentin@oato.inaf.it}
%\author{Mario G. Lattanzi}
%\affil{INAF - Osservatorio Astrofisico di Torino, Strada Osservatorio 20, 10025 Pino Torinese, TO, Italy; \\
%Shanghai Astronomical Observatory, Chinese Academy of Sciences, 80 Nandan Road, 200030 Shanghai, China}
%\email{lattanzi@oato.inaf.it}
%%\altaffiltext{2}{Shanghai Astronomical Observatory, Chinese Academy of Sciences, 80 Nandan Road, 200030 Shanghai, China}
%\author{Alessandro Spagna}
%\affil{INAF - Osservatorio Astrofisico di Torino, Strada Osservatorio 20, 10025 Pino Torinese, TO, Italy}
%\email{spagna@oato.inaf.it}
%\author{Anna Curir}
%\affil{IINAF - Osservatorio Astrofisico di Torino, Strada Osservatorio 20, 10025 Pino Torinese, TO, Italy}
%\email{curir@oato.inaf.it}

\begin{abstract}
We explore the vicinity of the Milky Way through the use of spectro-photometric data from the Sloan Digital Sky Survey and high-quality proper motions derived from multi-epoch positions extracted from the Guide Star Catalogue II database. In order to identify and characterise streams as relics of the Milky Way formation, we start with classifying, select, and study $2417$ subdwarfs 
with $\rm{[Fe/H] < -1.5}$
up to $3$~kpc away from the Sun as tracers of the local halo system. Then, through phase-space analysis, we find statistical evidence of five discrete kinematic overdensities 
among $67$ of the fastest-moving stars, 
and compare them to high-resolution N-body simulations of the interaction between a Milky-Way like galaxy and orbiting dwarf galaxies with 
four representative cases of merging histories. \\
The observed overdensities can be interpreted as fossil substructures consisting of streamers torn from their progenitors; 
such progenitors appear to be satellites on prograde and retrograde orbits on different inclinations. 
In particular, of the five detected overdensities, two appear to be associated, yelding twenty-one additional main-sequence 
members, with the stream of Helmi et al. (1999) that our analysis 
confirms on a high inclination prograde orbit. 
The three newly identified kinematic groups 
could be associated with the retrograde streams detected 
by Dinescu (2002) and Kepley et al. (2007); 
whatever their origin, the progenitor(s) would be on retrograde orbit(s) and inclination(s) 
within the range $10^{\circ} \div 60^{\circ}$. \\
Finally, we use our simulations to investigate the impact of observational errors 
and compare the current picture to the promising prospect of highly improved data expected from the Gaia mission.
\end{abstract}

\keywords{Galaxy: formation --- Galaxy: halo --- Galaxy: kinematics and dynamics}

%________________________________________________________________

\section{Introduction}\label{sec:1}

The formation and evolution of galaxies is one of the outstanding
problems in astrophysics, one which can be profitably engaged directly
through detailed study of our own Galaxy, the Milky Way 
\citep[e.g.,][]{FreemanBland-Hawthorn,HelmiRev}. 

In the context of hierarchical structure formation, galaxies such as the
Milky Way grow by mergers and accretion of smaller systems, perhaps
similar to what are now observed as dwarf galaxies. These satellite
galaxies -- torn apart by the tidal gravitational field of the parent
galaxy -- are progressively disrupted, giving rise to trails of stellar
debris streams along their orbits, spatial signatures that eventually disappear
due to dynamical mixing. After the accretion era ends, a spheroidal
halo-like component is left from their collective assembly 
\citep[e.g.,][]{SZ,BullockJohnston,Abadi,Moore,Sales,DeLucia}.  

Of all the Galactic components, it is indeed the stellar halo 
that offers the best opportunity for probing details of the merging history
of the Milky Way \citep[see, e.g.,][]{HelmiRev}. 
Past explorations have demonstrated that there is %the real 
a concrete possibility to identify groups of halo stars that originate from
common progenitor satellites \citep[][]{Eggen,Ibata94,Majewski,Helmi,ChibaBeers,
Dinescu,Ibata03,Kepley,Klement,Morrison, Schlaufman09,Smith2009, Duffau}.

Simulations show that a Milky-Way mass galaxy
within a $\rm{\Lambda}$CDM universe will have halo stars 
associated with substructures and streams 
\citep[e.g.,][]{Johnston98, Harding, Starkenburg, Helmi11, Gomez}. 
These substructures, much like those  
seen in the halo system of the Milky Way, are sensitive to recent (within the last $8$~Gyr)
merging events, and are more prominent in the outer region of the halo 
(galactocentric radii beyond $15-20$~kpc),
whereas the inner-halo region appears significantly smoother. 

Based on data from the SEGUE spectroscopic survey, 
\citet{Schlaufman09} found that metal-poor main-sequence turnoff stars in the
inner-halo region of the Milky Way (within $\sim 20$~kpc from the Sun) exhibit clear evidence for radial
velocity clustering on small spatial scales (they refer to these as
ECHOS, for Elements of Cold HalO Substructure). They estimated that
about $10\%$ of the inner-halo turnoff stars belong to ECHOS, and inferred
the existence of about $1000$ ECHOS in the entire inner halo volume.
\citet{Schlaufman11} suggest that the most likely progenitors of
ECHOS are dwarf spheroidal galaxies with masses on the order of
$10^{9}$~M$_{\sun}$.

In the Solar Neighbourhood, up to $1-2$~kpc of the Sun, 
stellar streams have also 
been discovered as overdensities in the 
phase-space distribution of stars, integrals of motion and action-angle variables 
\citep[see][for a review]{KlementRev}.
Prominent examples are the two stellar debris streams in the halo population passing close to the Sun 
detected by \citet{Helmi} 
when combining high-quality HIPPARCOS proper motions with 
ground-based observations. 
Formed via destruction of a satellite whose debris now occupy the
inner halo region with no apparent spatial structure, 
these streamers retain very similar velocities 
and are seen as clumps in angular momentum space 
where stars from a common progenitor appear   
rather confined \citep{HelmideZeeuw}.

Besides the Helmi stream, $\omega$ Centauri \citep[][]{Dinescu,Majewski2012}, 
the Kapteyn and Arcturus \citep[e.g.,][]{Eggen71} streams, 
\citet[][]{KlementRev} lists a few other halo substructures found in the solar neighborhood: 
these are still small numbers compared to the few hundred streams expected 
\citep[i.e. $300$-$500$,][]{HelmiWhite, Gould}.
Actually, recovering fossil structures in the inner halo is 
considerably more difficult, as strong phase-mixing takes place. This
degeneracy can only be broken with 6D (phase-space) or 7D (including
abundances) information achievable by integrating astrometry,
photometry, and spectroscopy. 

The SDSS~-~GSC~II Kinematic Survey (from now on SGKS) 
we exploit here was produced to serve this task \citep[see][]{Spagna10}.
In the future, new ground- and space-based surveys such as 
Gaia \citep[e.g.,][]{Perryman,Turon}, 
Gaia ESO Survey \citep[GES;][]{Gilmore}, and 
LAMOST \citep[][]{LAMOST} 
will provide high-precision data 
that will usher us in a new era of Milky Way studies.

In Sect.~\ref{sec:2}, we introduce the data
used to isolate a sample of nearby halo subdwarfs from the SGKS 
catalogue. The kinematic and orbital properties of the local halo
subdwarf population are discussed in Sect.~\ref{sec:3}, where we present
algorithms to search for kinematic substructures, recovering known
streams \citep[][]{Helmi, Dinescu, Kepley}, as well as 
new kinematic overdensities. 
In Sect.~\ref{sec:4}, we present the
high-resolution N-body numerical simulations of four minor mergers used
to study galaxy interactions and the properties of accretion events in
the vicinity of the Sun. 
In Sect.~\ref{sec:5} we investigate the impact of
observational errors resulting from current ground-based data and 
from high accurate data expected from the Gaia satellite. 
Finally, in Sect.~\ref{sec:6}, we compare observations
to these numerical simulations and infer the nature of 
the detected fast moving groups.

%________________________________________________________________ 

\begin{deluxetable*}{lcccccccccc}
\tablewidth{0pt}
\tablecaption{\label{table:1} Halo velocity parameters.
	}
\tablehead{
\colhead{$\langle U\rangle$}  & \colhead{$\langle V+220\rangle$}  & \colhead{$\langle W\rangle$}  & 
\colhead{$\sigma_U$}  &  \colhead{$\sigma_V$}  & \colhead{$\sigma_W$}  &
\colhead{$\rho_{UV}$}  &  \colhead{$\rho_{UW}$}  & \colhead{$\rho_{VW}$}\\
$\rm{(km~s^{-1})}$ & $\rm{(km~s^{-1})}$ & $\rm{(km~s^{-1})}$ & $\rm{(km~s^{-1})}$ & $\rm{(km~s^{-1})}$ & $\rm{(km~s^{-1})}$
}
\startdata
$15\pm 2$&$25\pm 2$&$-4\pm 2$ & $126\pm 1$ & $100\pm 1$ & $91\pm 1$ & $-0.09\pm 0.02$ & $-0.18\pm 0.02$ & $0.05\pm 0.02$\\
\enddata
\tablecomments{The Milky Way halo velocity parameters as determined from our selected sample of 2417 FGK subdwarfs. 
	The table lists mean velocities, dispersions, and corresponding correlation coefficients 
	in Galactic coordinates. Noticeable is the correlation between $U$ and $W$ (see text).}
%\tablenotetext{}{}
\end{deluxetable*}

\section{The SDSS~-~GSC~II Kinematic Survey (SGKS)}\label{sec:2}

This study is based on a new kinematic catalogue, derived by assembling
spectro-photometric stellar data from the Seventh Data Release of the
Sloan Digital Sky Survey \citep[SDSS DR7;][]{SDSS7}, which
included data from the Sloan Extension for Galactic Understanding and
Exploration \citep[SEGUE;][]{Yanny}, supplemented by astrometric
parameters extracted from the database used for the construction of the
Second Guide Star Catalogue \citep[GSC II;][]{Lasker}. 
This SDSS~-~GSC~II catalogue contains positions, proper motions, 
classification, and $ugriz$ photometry for $77$ million sources 
down to $r \sim 20$, over $9000$ square degrees. 

Proper motions are computed by combining multi-epoch positions from SDSS
DR7 and the GSC II database; typically, $5-10$ observations are available
for each source, spanning up to $50$ years.  
The typical formal 
errors on those proper motions are in the range 
$2-3~{\rm mas~yr^{-1}}$ per coordinate for $16<r<18.5$,
comparable with the internal precision of the 
SDSS proper motions computed by \citet{Munn}.
Although much of the photographic material (Schmidt plates) 
used to derive first epoch information is in common with \citet{Munn}, 
plate digitisation and measurement processes, and the calibration methods 
that led to the first epoch positions were somewhat different, 
with the Munn et al.' data coming from the USNO-B project. 
Of particular relevance is the 
minimisation of systematic errors that can affect proper motion accuracy, 
a true driver in analysis like those conducted in this study. 
An accurate validation of our proper motions was discussed  
in \citet{SpagnaL}.

Radial velocities and astrophysical parameters 
are available for about $151\,000$ 
sources cross-matched with the SDSS spectroscopic catalogue 
Typical accuracy are of $5-10~{\rm km~s^{-1}}$ in line of sight velocity, 
$250$~K in effective temperature, ${\rm T_{eff}}$, 
$0.25$~dex in surface gravity, $\log~g$,
and $0.20$~dex in metallicity, ${\rm[Fe/H]}$, 
as estimated within the SEGUE Spectral Parameter Pipeline 
\citep[SSPP; i. e., ][]{ReFiorentin07,LeeSSPP1,LeeSSPP2,SSPP3}.
We specify that the sample includes only objects with no problems 
related to the spectrum, and classified without any cautionary flag 
by the SSPP.
In case of multiple spectra, we take the spectrum with the highest
signal-to-noise ratio.

From the SDSS~-~GSC~II catalogue, we select as tracers sources with 
$4500~{\rm K}<{\rm T_{eff}} < 7500$~K and $\log~g > 3.5$, 
corresponding to FGK dwarfs. 

The observed magnitudes are corrected for interstellar absorption 
via the extinction maps of \citet{Schlegel} based on the
$6.1'$ resolution COBE/DIRBE dust map, 
that we preferred to the more recent, but of inferior resolution ($7'$-$14'$), 
reddening maps published by \citet{Schlafly2014}. 
Then, we transformed the $E(B-V)$ to the SDSS photometric system by adopting the 
extinction ratio $A_r/A_V=0.875$ 
\citep[from Table 1 of ][]{Girardi}, that is appropriate for our FGK dwarf sample.

Photometric distances good to $\sigma_d/d\sim 20\%$ 
are computed by means of the photometric parallax relation 
established for FGK main-sequence stars by \citet{Ivezic}.
Here, the metallicity-dependent absolute magnitude relations, $M_r = f(g-i, {\rm[Fe/H]})$, 
use the spectroscopic ${\rm[Fe/H]}$ instead of the photometric metallicity adopted by \citet{Ivezic}. 
We also apply the additional colour thresholds from \citet{Klement} in order to remove turn-off stars, 
whose estimated $M_r$ may be affected by residual systematic errors.

Galactic space-velocity components\footnote{
Throughout this work, $U$, $V$, and $W$ indicate Galactic velocity components 
relative to the Local Standard of Rest and follow the convention with 
$U$ positive toward the Galactic center, 
$V$ positive in the direction of Galactic rotation, and 
$W>0$ towards the
North Galactic Pole.
} 
are estimated under the assumption that 
the Sun is at a distance of $8$~kpc from the centre of the Milky Way, 
the Local Standard of Rest (LSR) rotates at $220~\rm{km~s^{-1}}$ about the Galactic center, 
and the peculiar velocity of the Sun relative to the LSR is 
$(U,V,W)_{\sun}=(10.00, 5.25, 7.17)~\rm{km~s^{-1}}$ \citep{DehnenBinney}.

Finally, in order to 
minimise the effect of outliers (e.g. mismatches, blends and sources with low S/N) and therefore 
obtain a sample with accurate distance and kinematics suitable for our stellar stream search,
we impose a threshold on proper-motion errors ($<10~\rm{mas~yr^{-1}}$ per component), 
constrain magnitudes to the range $13.5 < g < 20.5$,
limit the errors on the derived velocity components to better than $50~\rm{km~s^{-1}}$, 
and remove total space velocities above $600~\rm{km~s^{-1}}$.
These are the properties of the $24\,634$ stars listed in 
the SGKS catalogue.

%________________________________________________________________ 

\section{Data analysis}\label{sec:3}

Among the full sample of FGK dwarfs from the SGKS 
catalogue, we have selected specific sub-samples of tracers of the
Galactic halo population in the inner-halo region, and analysed their 
phase-space distribution.

Here, we focus on a sample of $2417$ metal-poor stars
($\rm{[Fe/H]<-1.5}$) outside the Galactic plane ($|z|>1$~kpc),
and located within $3$~kpc of the Sun. 
Within this volume, the selected sample has 
median errors on $U$, $V$, and $W$ of 
$12$, $13$, and $9~\rm{km~s^{-1}}$, respectively; 
this results in errors in the velocity difference between stellar pairs 
not exceeding $\sim 20~\rm{km~s^{-1}}$. 
Such a value is  
suited for careful investigations of substructure, as 
the kinematic analysis presented below will show.\\

%________________________________________________________________ 

   \begin{figure*}
   \centering
   \includegraphics[width=\linewidth]{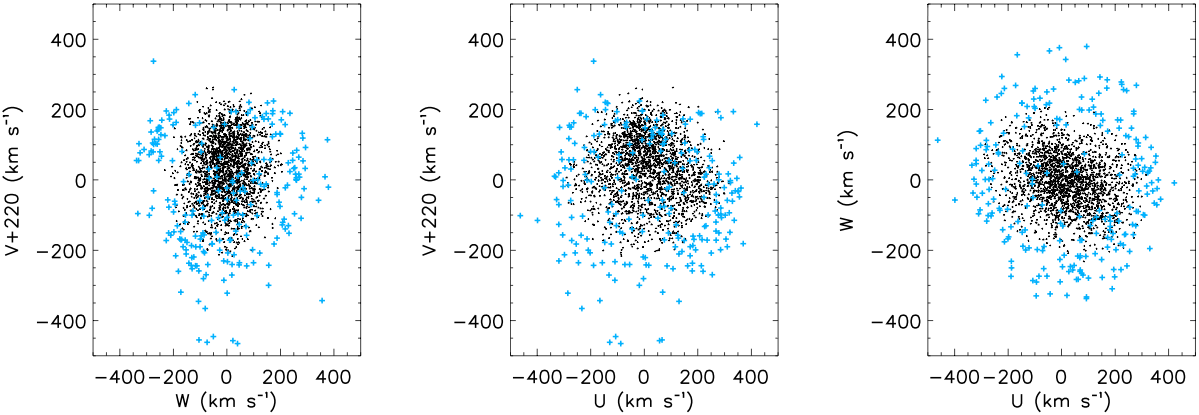}
   \caption{Distribution of nearby halo stars in velocity space for our selected sample of 
	$2417$ FGK subdwarfs, with 
	$\rm{[Fe/H] < -1.5}$ and $|z|>1$~kpc within $3$~kpc of the Sun. 
	The $10\%$ fastest-moving stars ($242$) are marked as crosses.
   }
   \label{PRF:fig1}
   \end{figure*}

\subsection{Local halo velocity distribution}\label{sec:3.1}

From the selected sample we measure the mean velocities
$(\langle U\rangle ,\langle V+220\rangle , \langle W\rangle)$,  
the velocity ellipsoid $(\sigma_U,\sigma_V,\sigma_W)$, and
the correlations among velocity components ($\rho_{UV}, \rho_{UW}, \rho_{VW})$ as 
reported in Table~\ref{table:1}.

The kinematic properties of the selected tracers are 
representative of the halo population in the vicinity of
the Sun \citep[e.g.,][]{ChibaBeers}.

The significant correlation $\rho_{UW}=-0.18\pm 0.02$ between the radial and vertical velocity components
indicates a tilt of the velocity ellipsoid (Fig.~\ref{PRF:fig1}, right panel). 

Using the tilt formula \citep[see, e.g.,][]{BinneyMerrifield}
\begin{equation}
\tan {2\delta_{UW}}= \frac{2~\sigma^2_{UW}}{\sigma^2_{U}-\sigma^2_{W}}= \frac{2~\rho_{UW}~\sigma_{U}\sigma_{W}}{\sigma^2_{U}-\sigma^2_{W}},
\label{e0}
\end{equation}
and the values in Table~\ref{table:1} for the correlation coefficient and velocity dispersions along the $U$ and $W$ axes, 
we derive a tilt angle of $\delta_{UW} = -14.5^\circ \pm 1.4^\circ$, 
revealing that the $(U, W)$ distribution points toward the Galactic center.

In fact, for our halo sample of 2417 FGK subdwarfs with 
$\langle z\rangle \approx $ 1.2 kpc and $\langle R\rangle \approx 8.3$~kpc, 
we estimate a mean position angle 
$\langle \tan^{-1}(z / R)\rangle~\approx 8.3^\circ$.
This result is fairly 
consistent with the tilting effects on the velocity ellipsoids due to 
the gravitational potential produced by the stellar disk and dark matter halo \citep[][and references therein]{Bond}. 
We also measure smaller but stastistically significant correlations in $(U,V)$ and $(V,W)$ velocity-planes. 

In the following, we look for halo streamers in the high velocity tail of the $(U,V,W)$ velocity 
distribution, where kinematic substructures are more easily detected.
In order to select high velocity stars, we model the velocity distribution as a tilted Schwarzschild ellipsoid:

\begin{equation}
\begin{array}{lll}
f(U,V,W)&=& {\rm const}\, \cdot  e^{-\frac{1}{2}E(U,V,W)}
\end{array}
\label{e1}
\end{equation}
where $E$ is the velocity function defined by: 
\begin{equation}
\begin{array}{lll}
&&\lefteqn{E(U,V,W)= }\\
&&\\
&&\frac{R_{UU}}{R}\left(\frac{U-\langle U\rangle}{\sigma_U}\right)^2+\frac{R_{VV}}{R}\left(\frac{V-\langle V\rangle}{\sigma_V}\right)^2+\frac{R_{WW}}{R}\left(\frac{W-\langle W\rangle}{\sigma_W}\right)^2+\\
&&\\
&&2\frac{R_{UV}}{R}\left(\frac{U-\langle U\rangle}{\sigma_U}\right)\left(\frac{V-\langle V\rangle}{\sigma_V}\right)+\\
&&\\
&&2\frac{R_{VW}}{R}\left(\frac{V-\langle V\rangle}{\sigma_V}\right)\left(\frac{W-\langle W\rangle}{\sigma_W}\right)+\\
&&\\
&&2\frac{R_{UW}}{R}\left(\frac{U-\langle U\rangle}{\sigma_U}\right)\left(\frac{W-\langle W\rangle}{\sigma_W}\right).
\end{array}
\label{e2}
\end{equation}

Here, $R$ represents the determinant of the symmetrical matrix $\mathcal{R}$ 
of the correlation coefficients $\rho_{ij}=R_{ij} / R$ (for $i,j=U,V+220,W$), 
and $R_{ij}$ designate the cofactor of the corresponding correlation element in $\mathcal{R}$ 
\citep[e.g.,][]{Trumpler}.

Figure~\ref{PRF:fig1} shows the kinematic distribution for the individual components, $(U, V+220, W)$, of
the space-velocity vector for the full sample of $2417$ selected halo stars; 
the $242$ objects comprising the sample of the $10\%$ highest velocity tail 
are represented with crosses.

As expected, the overall velocity distribution is relatively smooth, because of the strong phase-mixing that
takes place in the inner-halo region, and slowly prograde \citep[e.g.,][]{HelmiRev}.\\
However,  
as their motions (in direction and speed) are well separated from 
those of the other nearby subdwarfs, we intend to study the degree of clumpiness 
of the $10\%$ fastest-moving objects.
The case study is that, of all the objects passing within a few kiloparsecs of the Sun, 
some are part of a diffuse local stellar halo, 
while some could be debris of accretion events and remnants from the outer-halo
population currently in the Solar Neighbourhood.

Before starting to look for kinematic substructures, 
we check for thick disk stars that possibly contaminate our halo tracers. 
Here, we applied to kinematic method described in \citet{Spagna04} and estimate 
the fraction of subdwarfs that is consistent with the 3D velocity distribution 
of the thick disk population.
By assuming a velocity ellipsoid, as estimated by \citet[][]{Pasetto},
and a rotation velocity $V_{\phi} = 150$ km s$^{-1}$, as measured by \citet{SpagnaL} 
for metal-poor thick disk stars with $\rm{[Fe/H]}\simeq -1$ dex, we found at a $2\sigma$ (i.e. $87\%$) confidence 
level a $\sim 10\%$ maximum contamination of thick disk in the whole sample of 2417 halo tracers.
Instead, no contaminant is expected among the subsample of the $10\%$ fastest objects. 

We use the samples described above to detect and subsequently identify 
kinematic halo substructures in the Solar Neighbourhood as
groups of stars moving with similar velocities and directions. 
Detection is accomplished by performing 
a statistical test based on individual kinematics   
aimed at quantifying possible deviations from a smooth distribution 
of the background halo; 
cluster analysis in velocity space is then applied 
for final confirmation of the substructures.

%________________________________________________________________

\subsection{The two-point correlation function: finding the clumps}\label{sec:3.2}

The amount of kinematic substructures that cosmology might leave in the volume is
quantified by means of the cumulative two-point correlation function, $\xi({\bf v})$, 
on the paired velocity difference ${\bf v}=\vert{\bf v}_i-{\bf v}_j\vert$ 
that measures the excess in the number of stellar pairs moving 
within a given velocity difference when compared to a representative 
random smooth sample \citep[cfr.][for more details]{ReFiorentin05}.
Here the random points were drawn from a multivariate distribution obtained 
from the observed data set by random permutations of the order 
of the velocity components $V+220$ and $W$, after fixing $U$; 
finally, the actual random pairs are obtained after averaging over ten independent realisations.

This function is computed over the full sample of $2417$ halo stars, and
separately for the sub-sample of the $242$ fastest-moving stars,
corresponding to the $10\%$ high-velocity tail. 

A statistical excess of stars  with small pairwise velocity differences indicates the
presence of likely streamers made of objects with coherent kinematics. 

\begin{figure}[hb!]
   \centering
   \includegraphics[width=\linewidth]{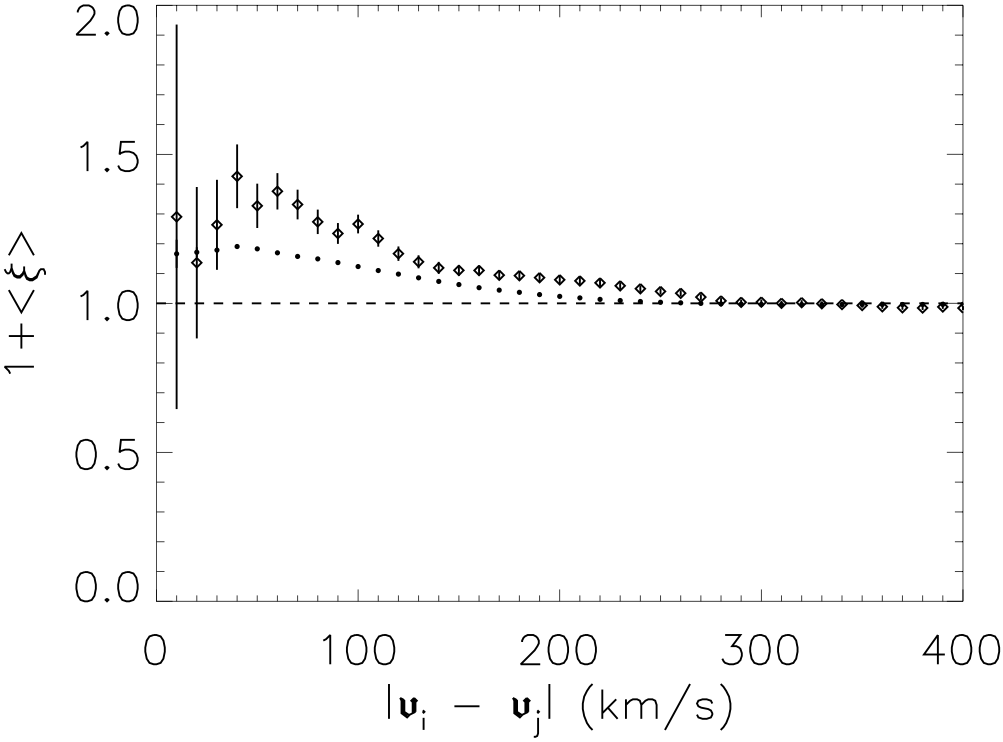}
   \caption{Cumulative velocity  
	correlation function for the full sample of halo stars (dots),
      and the $10\%$ fastest-moving subset (diamonds) shown in Fig.~\ref{PRF:fig1}.
	The error bars are derived from Poisson's statistics of the counts.
              }
         \label{PRF:fig2}
   \end{figure}

Figure~\ref{PRF:fig2} shows, using bins of $10~\rm{km\, s^{-1}}$ width\footnote{
We fixed the bin width following the rule that the interval sampled is divided into 
as many bins as the square rooth of the sample size, 
in our case $\sim 400~\rm{km\, s^{-1}}/\sqrt{2417} \sim 10~\rm{km\, s^{-1}}$.}
%$10~\rm{km\, s^{-1}}$,
the two-point correlation function $\xi({\bf v})$ 
for the full sample of $2417$ halo stars (dots) and 
for the subset of the $10\%$ fastest-moving stars (diamonds). 
While weak for the full sample,
there is a statistically significant signal ($SNR > 4$)
for the subset of the fastest stars, that peaks at $40~\rm{km\, s^{-1}}$: 
the excess of pairs of stars with similar velocities
is very noticeable, and is a direct indication of the presence of
kinematical clumps.

\begin{figure*}[ht!]                                                                                                         
   \centering
   \includegraphics[width=\textwidth]{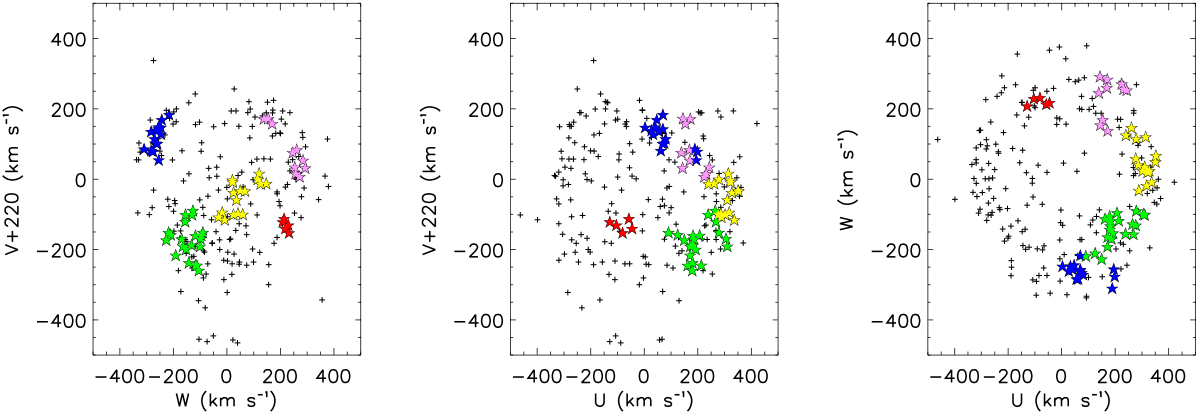}
   \caption{Distribution of the high-velocity tail 
	from our selected sample with
     $\rm{[Fe/H] < -1.5}$ and $\rm{|z|> 1}$~kpc within $3$~kpc of the Sun (see Fig.~\ref{PRF:fig1}).     
     Shown are $242$ objects, the $10\%$ fastest-moving stars.  
	Among them, star symbols identify 
     the $67$ sources with pairwise velocity differences below $40~\rm{km~s^{-1}}$;
	the stars belonging to isolated pairs have been excluded.
     Different colours are used to indicate stars associated with the five
     clumps recovered by the clustering analysis. 
     }
   \label{PRF:fig3}
   \end{figure*}

In the following, among the sample of the $242$ fastest stars,  
we focus on the objects with paired velocity differences less than $40~\rm{km\, s^{-1}}$, 
which yield the statistically significant signal seen in Fig.~\ref{PRF:fig2}.
In addition, we exclude {\it isolated} pairs, i.e., ``groups" with only two objects. 
This further selection certainly reduces the number of detected members, 
however it makes the following analysis more robust by decreasing the contamination 
of false positives.
The final sample is made of $67$ stars.\\

%________________________________________________________________

\subsection{Clustering analysis: assigning membership}\label{sec:3.3}

In order to classify these $67$ objects, 
we perform $K-$medoids clustering\footnote{We used the 
implementation of the $K-$medoids clustering developed as part of the {\it R Project for Statistical Computing}: 
www.r-project.org} 
in the 3D velocity space 
that defines the number of kinematic substructures and their members.
This unsupervised learning algorithm is able to group data
into a pre-specified number of clusters that minimises the RMS of the distance 
(in velocity space) to the center of each cluster.
 
The original data set is
initially partitioned into clusters around $K$ data points referred to as the medoids,
then an iterative scheme (PAM, for Partitioning Around Medoids) is applied to locate the medoids that 
achieve the lowest configuration ``cost". 
The algorithm employed by PAM,  
similar to the $K$-means clustering
algorithm, is more robust to outliers and 
obtains a unique partitioning of the data without the need for explicit multiple starting
points for the proposed clusters 
\citep[see, e.g., ][]{Kaufmann,Hastie}.

There is no general theoretical solution for finding the optimal number of 
clusters for any given data set. 
Increasing $K$ results in the error function values formally much smaller, 
but this increases the risk of overfitting.
In order to keep the final identification safe and simple,
we compared the results of runs with different 
$K$ classes, and the best $K$ resulted following visual inspection of the generated distribution. 
The solution adopted here is with $K=5$: 
for $K<5$ the clusters returned by the algorithm would contain a mixture of the natural underlying groups 
(e.g., prograde and retrograde members in the same kinematic clump); 
for $K>5$ natural groups further partition into ``artificial" subgroups.

The five kinematic substructures detected are visualised in Fig.~\ref{PRF:fig3} 
with different colours. 

The individual kinematic properties of the $67$ stars 
belonging to the five kinematics groups are listed in Table~\ref{table:2}.

Other methods have been utilised to isolate groups of stars in the halo like, e.g., 
the interesting approach specifically developed  
for the Virgo stellar stream by \citet{Duffau}. 
On the other hand, the fact that we have the complete set of 
3D kinematical data and that the whole sample is confined within $3$~kpc from the Sun 
(i.e., the distance segregation is implicitly implemented in our sample) 
suggests the direct use of a classical clustering algorithm like PAM as the method of choice.

%________________________________________________________________

\subsection{Angular momentum and orbital properties}\label{sec:3.4}

\begin{figure*}[ht!]
   \centering
   \includegraphics[width=\linewidth]{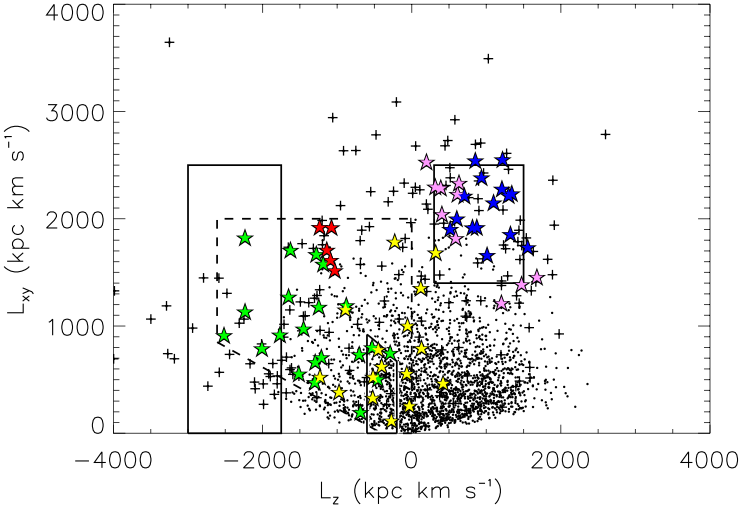}
   \caption{Distribution of the selected sample of $2417$ FGK subdwarfs within $3$~kpc of the Sun 
	in the space of adiabatic invariants.
	Cross symbols indicate the $10\%$ fastest-moving objects. 
     Among them, star symbols identify         
     the $67$ sources with paired velocity differences below $40~\rm{km~s^{-1}}$. 
     As in Fig.~\ref{PRF:fig3}, different colours are used to indicate stars associated with the five
     clumps recovered by the clustering analysis in velocity space. 
     At $L_z>0$, the solid box shows the locus of the halo stream discovered by \citet{Helmi}:
     the kinematic substructures, pink and blue stars (i.e., Groups 1 and 2 in Table~\ref{table:2}), 
     on prograde orbits indeed cover the same region.
     At $L_z<0$, the solid box at $L_z \sim -2375$ shows the locus of the 
	substructure detected by \citet{Kepley} as retrograde outliers, 
	while the solid contour at $L_z \sim -400$ identifies the $\omega$~Cen substructures 
	remapped from the $L_z$-$L$ region in \citet{Dinescu}.
	The area within the dashed line includes the kinematic Groups 3, 4, and 5 of Table~\ref{table:2}, 
	represented by the red, green, and yellow stars, respectively.
     }
   \label{PRF:fig4}
   \end{figure*}

The space of adiabatic invariants allows better identification of the
different possible merging events that might have given rise to the
observed substructures. Clumping should be even stronger, since all
stars originating from the same progenitor should have very similar
integrals of motion, resulting in a superposition of the corresponding
streams; that is, the initial clumping of satellites are present even
after the system has completely phase-mixed \citep[e.g.,][]{HelmideZeeuw}.

In this study we focus on the plane defined by the 
components of angular momentum\footnote{Remind that: 
$L_x=yv_z-zv_y$, $L_y=zv_x-xv_z$, $L_z= xv_y-yv_x$, and $L_{xy}=\sqrt{L_x^2+L_y^2}$. 
Here $v_x=-U$, $v_y=V+220$, and $v_z=W$.} 
in and out the plane of the Galaxy's disk, i.e. $L_{xy}$ and $L_{z}$ respectively.

Since for a local sample $(x,y,z)\sim (r_{\sun},0,0)$, 
we remark that 
$L_{xy}\sim r_{\sun} |v_z|$ is dominated by the velocity perpendicular to the plane
and $L_z\sim r_{\sun} v_y$  measures the amount of rotation of a given stellar  orbit.
Essentially, 
stars with high/low $L_{xy}$ are on high/low-inclination orbits; 
stars with $L_z < 0$ are on retrograde orbits and 
stars with $L_z > 0$ are on prograde orbits.  

Figure~\ref{PRF:fig4} shows the distribution of the selected sample
within $3$~kpc of the Sun in the angular momentum diagram $L_z$ versus
$L_{xy}$. As in Fig.~\ref{PRF:fig3}, the $10\%$ fastest-moving objects are
plotted as crosses, and with the star symbols we mark the group members 
identified in Sect.~\ref{sec:3.2}. 
Different colours indicate the stars associated with the different lumps recovered by the
cluster analysis in velocity space. 

The solid lines show the loci of the known kinematic
structures detected by 

\newpage

\begin{enumerate}
\item 
\citet{Helmi} at \\%and further enlarged by \citet{ReFiorentin05} at     
$300<L_{z}<1500~\rm{kpc~km~s^{-1}}$ and \\
$1400<L_{xy}<2500~\rm{kpc~km~s^{-1}}$, 
\item 
\citet{Kepley} at \\
$-3000<L_{z}<-1750~\rm{kpc~km~s^{-1}}$ and \\
$0<L_{xy}<2500~\rm{kpc~km~s^{-1}}$,

\item\citet{Dinescu} at\footnote{For the \citet{Dinescu} region, the curve delimiting the $L_{xy}$ upper part 
was derived by remapping the $L_z$-$L$ region shown in Fig. 4 of that paper;
therefore $671<L_{xy}^{lim}<922$.} \\
$-600<L_{z}< -200~\rm{kpc~km~s^{-1}}$ and \\
$0<L_{xy}<L_{xy}^{lim}~\rm{kpc~km~s^{-1}}.$ 
\end{enumerate}

\begin{deluxetable*}{llrrrrrr}
%\tabletypesize{\small}
%\tabletypesize{\scriptsize}
\tablewidth{0pt}
\tablecaption{\label{table:2} Main individual characteristics of the $67$ fastest-moving stars found members of 5 different kinematics groups.
	}
\tablehead{
\colhead{Group}           & \colhead{ID}      &
\colhead{$U$}          & \colhead{$V+220$}  &
\colhead{$W$}          & \colhead{$L_{xy}$}    &
\colhead{$L_z$}  & \colhead{$\rm{[Fe/H]}$}\\
      &      & $\rm{(km~s^{-1})}$ & $\rm{(km~s^{-1})}$ & $\rm{(km~s^{-1})}$ & $\rm{(kpc~km~s^{-1})}$& $\rm{(kpc~km~s^{-1})}$ & (dex)
%\colhead{ref}
}
%Group  & ID & $U$ & $V+220$& $W$ & $L_{xy}$ & $L_z$ & $\rm{[Fe/H]}$ \\
%      &      & $\rm{(km~s^{-1})}$ & $\rm{(km~s^{-1})}$ & $\rm{(km~s^{-1})}$ & $\rm{(kpc~km~s^{-1})}$& $\rm{(kpc~km~s^{-1})}$ & (dex)\\
\startdata
1$^a$ / pink &52209-0694-596 & $ 143\pm 14$ & $ 33\pm 12$ & $ 293\pm 6$ & $ 2289 $ & $ 389 $ & $ -2.01$ \\
 & 52721-1050-418 & $ 145\pm  7$ & $ 173\pm 6$ & $ 155\pm 4$ & $ 1390 $ & $ 1473 $ & $ -2.02$ \\
 & 53315-1907-393 & $ 170\pm 31$ & $  54\pm 31$ & $ 285\pm 16$ & $ 2232 $ & $  614 $ & $ -1.75$ \\
 & 53349-2066-511 & $ 154\pm 23$ & $ 158\pm 28$ & $ 169\pm 27$ & $ 1459 $ & $ 1678 $ & $ -1.85$ \\
 & 54175-2472-370 & $ 139\pm 13$ & $  76\pm 16$ & $ 246\pm 4$ & $ 2336 $ & $  634 $ & $ -1.75$ \\
 & 54574-2904-114 & $ 243\pm 9$ & $  35\pm 9$ & $ 254\pm 3$ & $ 2301 $ & $  316 $ & $ -1.94$ \\
 & 54577-2906-088 & $ 225\pm 12$ & $   8\pm 12$ & $ 272\pm 5$ & $ 2532 $ & $  198 $ & $ -2.16$ \\
 & 54621-2191-177 & $ 231\pm 11$ & $  16\pm 11$ & $ 257\pm 10$ & $ 2045 $ & $  405 $ & $ -1.81$ \\
 & 54624-2189-236 & $ 169\pm 17$ & $  84\pm 26$ & $ 261\pm 23$ & $ 1820 $ & $  590 $ & $ -1.73$ \\
 & 54629-2902-237 & $ 172\pm 7$ & $ 172\pm 9$ & $ 138\pm 6$ & $ 1216 $ & $ 1203 $ & $ -2.64$ \\
\tableline%\cutinhead{This is a cut-in head}
2$^a$ / blue & 52316-0559-336 & $ 189\pm 3$ & $ 87\pm 11$ & $ -310\pm 5$ & $ 2545 $ & $ 853 $ & $ -2.52 $ \\
 & 53084-1368-399 & $ 195\pm 5$ & $  55\pm 6$ & $ -255\pm 4$ & $ 1906 $ & $  517 $ & $ -2.16$ \\
 & 53242-1896-109$^c$ & $  41\pm 14$ & $ 149\pm 11$ & $ -256\pm 7$ & $ 2220 $ & $ 1299 $ & $ -1.66$ \\
 & 53262-1900-359 & $  62\pm 19$ & $  82\pm 21$ & $ -283\pm 17$ & $ 2215 $ & $  706 $ & $ -2.09$ \\
 & 53293-1906-633$^c$ & $  27\pm 6$ & $ 139\pm 7$ & $ -261\pm 5$ & $ 2281 $ & $ 1210 $ & $ -1.91$ \\
 & 53467-2110-134 & $  69\pm 10$ & $ 102\pm 11$ & $ -260\pm 3$ & $ 1921 $ & $  815 $ & $ -1.55$ \\
 & 53712-2314-639 & $  75\pm 5$ & $ 103\pm 6$ & $ -265\pm 5$ & $ 2387 $ & $  936 $ & $ -2.57$ \\
 & 53726-2306-188$^c$ & $  55\pm 9$ & $ 136\pm 10$ & $ -283\pm 7$ & $ 2555 $ & $ 1218 $ & $ -1.74$ \\
 & 53907-2209-540$^c$ & $  70\pm 13$ & $ 144\pm 12$ & $ -256\pm 13$ & $ 1663 $ & $ 1011 $ & $ -2.30$ \\
 & 54178-2452-540 & $  70\pm 6$ & $ 184\pm 7$ & $ -216\pm 3$ & $ 1735 $ & $ 1556 $ & $ -1.51$ \\
 & 54380-2323-448 & $  47\pm 15$ & $ 170\pm 13$ & $ -243\pm 13$ & $ 1858 $ & $ 1324 $ & $ -2.12$ \\
 & 54479-2867-531 & $  33\pm 14$ & $ 129\pm 11$ & $ -246\pm 12$ & $ 2153 $ & $ 1097 $ & $ -1.64$ \\
 & 54530-2889-458 & $   3\pm 4$ & $ 148\pm 9$ & $ -247\pm 4$ & $ 2238 $ & $ 1342 $ & $ -1.56$ \\
 & 54554-2918-615 & $ 198\pm 15$ & $  80\pm 15$ & $ -276\pm 3$ & $ 2004 $ & $  604 $ & $ -1.98$ \\
 & 54580-2905-169 & $  80\pm 10$ & $ 115\pm 11$ & $ -271\pm 4$ & $ 1923 $ & $  873 $ & $ -1.71$ \\
\tableline%\cutinhead{This is a cut-in head}
3$^b$ / red & 54179-2567-458 & $ -57\pm 12$ & $ -112\pm 16$ & $ 214\pm 8$ & $ 1922 $ & $ -1076 $ & $ -1.64$ \\
 & 54463-2856-571 & $ -129\pm 9$ & $ -121\pm 22$ & $ 209\pm 6$ & $ 1617 $ & $ -1089 $ & $ -2.09$ \\
 & 54551-2394-228 & $ -104\pm 10$ & $ -130\pm 14$ & $ 230\pm 8$ & $ 1924 $ & $ -1234 $ & $ -1.72$ \\
 & 54562-2920-596 & $  -81\pm 19$ & $ -152\pm 19$ & $ 232\pm 10$ & $ 1521 $ & $ -1032 $ & $ -2.36$ \\
 & 54569-2900-046 & $  -46\pm 7$ & $ -139\pm 8$ & $ 218\pm 3$ & $ 1712 $ & $ -1139 $ & $ -1.68$ \\
\tableline%\cutinhead{This is a cut-in head}
4$^b$ / green  & 52059-0597-072 & $ 125\pm 4$ & $ -157\pm 10$ & $ -210\pm 6$ & $ 1672 $ & $ -1280 $ & $ -2.09$ \\
 & 52338-0788-070 & $ 180\pm 9$ & $ -237\pm 11$ & $ -142\pm 9$ & $  921 $ & $ -1767 $ & $ -2.07$ \\
 & 52942-1509-488 & $ 179\pm 9$ & $ -259\pm 22$ & $ -105\pm 8$ & $ 1137 $ & $ -2235 $ & $ -1.64$ \\
 & 53710-2310-141 & $ 207\pm 39$ & $ -159\pm 16$ & $  -95\pm 20$ & $ 1195 $ & $  -877 $ & $ -1.60$ \\
 & 53762-2381-588 & $ 213\pm 16$ & $ -244\pm 29$ & $ -117\pm 14$ & $  916 $ & $ -2517 $ & $ -1.85$ \\
 & 53800-2383-625 & $ 172\pm 7$ & $ -190\pm 14$ & $ -164\pm 8$ & $ 1275 $ & $ -1652 $ & $ -1.69$ \\
 & 53823-2240-213 & $ 310\pm 22$ & $ -190\pm 26$ & $ -100\pm 4$ & $  557 $ & $ -1513 $ & $ -1.70$ \\
 & 53874-2173-414 & $ 279\pm 16$ & $ -151\pm 18$ & $  -89\pm 12$ & $  199 $ & $  -688 $ & $ -1.57$ \\
 & 54154-2701-341 & $ 171\pm 10$ & $ -216\pm 16$ & $ -191\pm 21$ & $ 1826 $ & $ -2236 $ & $ -1.74$ \\
 & 54169-2413-090 & $ 149\pm 12$ & $ -172\pm 12$ & $ -226\pm 7$ & $ 1711 $ & $ -1626 $ & $ -1.77$ \\
 & 54234-2663-321 & $  90\pm 11$ & $ -151\pm 13$ & $ -218\pm 3$ & $ 1581 $ & $ -1187 $ & $ -2.16$ \\
 & 54539-2894-314 & $ 162\pm 3$ & $ -244\pm 12$ & $ -112\pm 5$ & $  793 $ & $ -2008 $ & $ -2.12$ \\
 & 54539-2894-632 & $ 305\pm 6$ & $ -168\pm 6$ & $  -99\pm 3$ & $  478 $ & $ -1299 $ & $ -1.59$ \\
 & 54544-2459-072 & $ 239\pm 7$ & $ -100\pm 8$ & $ -154\pm 7$ & $  802 $ & $  -532 $ & $ -1.82$ \\
 & 54557-2177-009 & $ 277\pm 16$ & $ -100\pm 18$ & $ -130\pm 10$ & $  507 $ & $  -454 $ & $ -2.73$ \\
 & 54568-2899-316 & $ 182\pm 17$ & $ -160\pm 25$ & $ -117\pm 4$ & $  665 $ & $ -1294 $ & $ -1.92$ \\
 & 54595-2932-091 & $ 268\pm 17$ & $ -120\pm 15$ & $ -156\pm 7$ & $  737 $ & $  -704 $ & $ -1.63$ \\
 & 54597-2561-326 & $ 265\pm 15$ & $  -89\pm 3$ & $ -127\pm 4$ & $  750 $ & $  -289 $ & $ -1.96$ \\
 & 54616-2460-420 & $ 186\pm 8$ & $ -169\pm 9$ & $ -169\pm 9$ & $ 1180 $ & $ -1246 $ & $ -1.93$ \\
 & 54616-2460-616 & $ 192\pm 16$ & $ -199\pm 15$ & $ -152\pm 16$ & $  977 $ & $ -1450 $ & $ -1.98$ \\
 & 54631-2911-151 & $ 182\pm 11$ & $ -187\pm 11$ & $ -132\pm 8$ & $  710 $ & $ -1212 $ & $ -1.96$ \\
\tableline%\cutinhead{This is a cut-in head}
5$^b$ / yellow & 53035-1433-600 & $ 322\pm 10$ & $-8\pm 13$ & $ 22\pm 6$ & $ 1004 $ & $ -56 $ & $ -1.58$ \\
 & 53240-1894-079 & $ 277\pm 22$ & $ -99\pm 18$ & $  59\pm 18$ & $  120 $ & $ -273 $ & $ -2.10$ \\
 & 53315-1907-353 & $ 352\pm 35$ & $ -33\pm 35$ & $  68\pm 18$ & $  258 $ & $  -32 $ & $ -1.50$ \\
 & 53770-2387-010 & $ 322\pm 6$ & $  -2\pm 12$ & $  21\pm 4$ & $  559 $ & $  -66 $ & $ -1.67$ \\
 & 53848-2437-060 & $ 314\pm 13$ & $  17\pm 22$ & $ 121\pm 13$ & $ 1787 $ & $ -228 $ & $ -2.18$ \\
 & 53876-2134-516 & $ 318\pm 13$ & $ -94\pm 13$ & $ -17\pm 5$ & $  528 $ & $ -518 $ & $ -1.52$ \\
 & 53918-2539-196 & $ 240\pm 10$ & $ -13\pm 8$ & $ 126\pm 8$ & $ 1356 $ & $  121 $ & $ -1.93$ \\
 & 54082-2325-126 & $ 278\pm 23$ & $   0\pm 16$ & $ 115\pm 15$ & $  468 $ & $  420 $ & $ -1.61$ \\
 & 54156-2393-459 & $ 336\pm 11$ & $-115\pm 9$ & $  -8\pm 5$ & $  527 $ & $-1231 $ & $ -1.74$ \\
 & 54243-2176-476 & $ 350\pm 6$ & $ -32\pm 4$ & $  50\pm 4$ & $  795 $ & $  131 $ & $ -1.73$ \\
 & 54271-2449-590 & $ 261\pm 11$ & $ -12\pm 9$ & $ 147\pm 7$ & $ 1686 $ & $  321 $ & $ -2.21$ \\
 & 54368-2804-351 & $ 302\pm 15$ & $-100\pm 19$ & $  24\pm 19$ & $  332 $ & $ -523 $ & $ -1.67$ \\
 & 54536-2871-426 & $ 286\pm 21$ & $ -97\pm 40$ & $  42\pm 11$ & $ 1155 $ & $ -888 $ & $ -1.93$ \\
 & 54594-2965-227 & $ 288\pm 14$ & $-107\pm 16$ & $ -31\pm 6$ & $  387 $ & $ -974 $ & $ -1.64$ \\
 & 54594-2965-272 & $ 328\pm 14$ & $ -37\pm 12$ & $  25\pm 5$ & $  782 $ & $ -448 $ & $ -1.54$ \\
 & 54616-2929-342 & $ 310\pm 7$ & $ -58\pm 8$ & $  36\pm 2$ & $  625 $ & $ -403 $ & $ -1.60$ \\
%\cutinhead{This is a cut-in head}
%\sidehead{I'm a side head:}
\enddata
\tablecomments{Main individual characteristics of the $67$ selected fastest-moving
	stars with paired velocity differences less than $40\,\rm{km\,s^{-1}}$, 
	and belonging to 5 different kinematics groups.
	These are 25 subdwarfs (21 new) on 2 streamers (Group~1 and Group~2) both members of the stream known to \citet{Helmi} 
	originally made of red giants and RR Lyrae.  
	The remaining $42$ are subdwarfs members of three newly discovered kinematic groups (Group~3, 4, and 5);
	see text for their dynamical interpretation.}
\tablenotetext{a}{Subdwarf members associated with the \citet{Helmi} stream.}
\tablenotetext{b}{Subdwarf members of the newly discovered kinematic groups.}
\tablenotetext{c}{Subdwarf members classified by \citet{Klement}.}
%\tablenotetext{d}{Subdwarf members associated with $\omega$~Cen stream.}
%% You can append references to a table using the \tablerefs command.
%%\tablerefs{(1) Barbuy, Spite, \& Spite 1985; (2) Bond 1980; (3) Carbon et al. 1987}
\end{deluxetable*}

The most noticeable feature in Fig.~\ref{PRF:fig4}  
is certainly the kinematic group corresponding to the stream found by \citet{Helmi}. 
Here, we identify $25$ subdwarfs, including $4$ stars already detected by \citet{Klement}, and $21$ new members.
By inspection of Fig.~\ref{PRF:fig3} we notice that 
the $10$ members belonging to Group~1 (pink star symbols) run along near-parallel orbits and cross the Milky Way's disk at 
high speed from South to North, and  
the $15$ objects in Group~2 (blue star symbols) 
cross the Milky Way's disk at similar speed and angle, but from North to South.

The three remaining lumps of fast-moving stars (red Group~3, green Group~4, and yellow Group 5) 
appear on the retrograde side of Fig.~\ref{PRF:fig4}.
The pentagonal box confined by the dashed line includes most of the members of 
Group~3 ($5$ stars), Group~4 ($21$ stars), and Group 5 ($16$ stars). 

These groups, and in particular the small Group~3, do not appear to be easily  associated with known streams 
and, in Sect.~\ref{sec:5}, 
we discuss the possibility that all these stars come from a common progenitor or from three different merging events.

Anyhow, we note that Group~4 might be the parent populations of 
the counter-rotating ``outliers'', with $V_{\phi} < -250~\rm{km~s^{-1}}$, found by \citet{Kepley}, 
while the slightly retrograde Group~5 ($V_\phi \approx -50~\rm{km~s^{-1}}$) 
seems to be related to the kinematic structure found by \citet{Dinescu}, 
and confirmed by both \citet{Meza} and \citet{Majewski2012}. 
These authors have also discussed the possibility that such a stream is formed by the tidal debris of $\omega$~Cen in the solar neighborhood, even though \citet{Navarrete} have recently ruled out this hypothesis 
after detailed analysis of the chemical abundance of this group 
with respect to the well-known peculiar properties of $\omega$~Cen.

%________________________________________________________________

\section{Simulations}\label{sec:4}

We explore a simulated inner halo based on a set of {\it four} 
high-resolution numerical N-body simulations of minor mergers. 
We analyse the kinematics and orbital properties of these simulations in order to 
investigate and characterise detectable signatures.  

   \begin{figure*}
   \centering
   \includegraphics[angle=0, width=\linewidth]{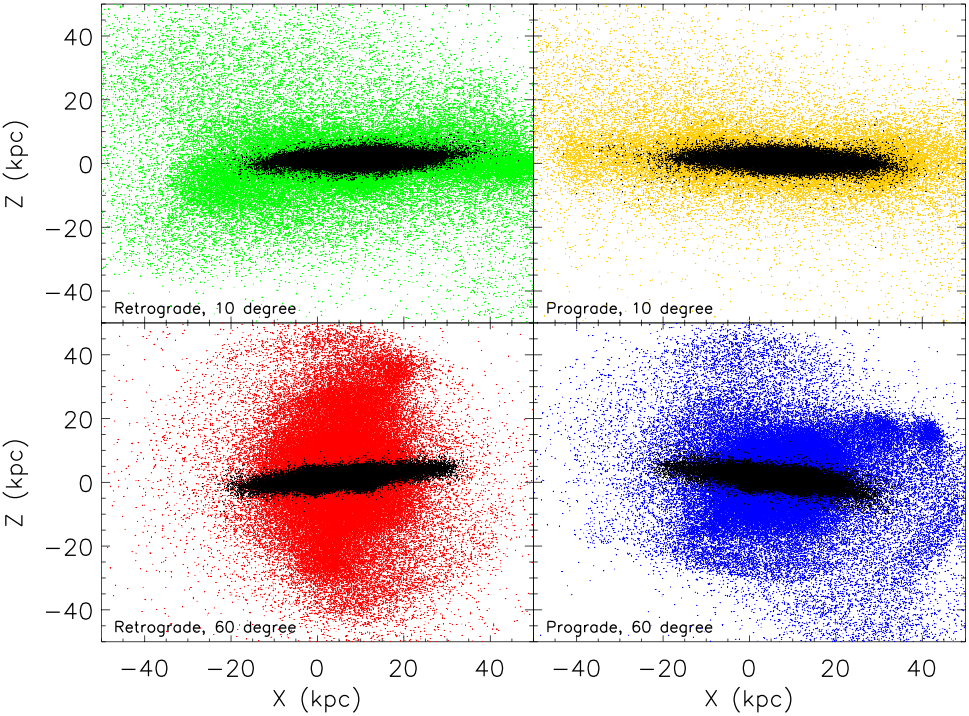}
   \caption{Final configurations in the $(x,z)$ plane of four minor merger events: 
     depicted are the morphologies of the stellar distribution, 
     i.e. the host disk (black) and the satellite bulge (colour), 
     at the final time $T=4.63$~Gyr of the simulations. 
     Shown are the cases of low-inclination ($10^{\circ}$ tilt) retrograde (top left) / prograde (top right) orbit, 
     and of high-inclination ($60^{\circ}$ tilt) retrograde (bottom left) / prograde (bottom right) orbit. 
     The panels only display a randomly selected $10\%$ subset of the total particles utilised.
              }
         \label{PRF:fig5}
   \end{figure*}

It is useful to point out  that these simulations are not an attempt to ``fit'' the observations,
but they represent four 
merging events that we assume as representative, in terms of 
inclination and rotation, of the initial orbits of the satellites.
In particular, these choices allow us to analyse the two cases suffering maximum and minimum dynamical friction. 

\subsection{N-body simulations}\label{sec:4.1}

We use a set of high-resolution numerical N-body simulations which
simulate minor mergers of prograde and retrograde orbiting satellite
halos within a dark matter main halo \citep{Murante}. The main
DM halo, which contains a stellar, rotating exponential disk, has a NFW
radial density profile \citep[]{Navarro}, with a mass
($M_{200}=10^{12}$~M$_{\sun}$), radius ($R_{200}=165$~kpc), and concentration
($C_{200}=7.5$), appropriate for a Milky Way-like DM halo at redshift $z = 0$; 
the spin parameter is set\footnote{
The cases $\lambda=0$ and $\lambda=1$ were both studied at lower resolution and the results compared; 
the differences are such that they have no bearing on the results presented in this paper.
} 
to $\lambda=1$. 
The satellite is
represented by a secondary DM halo containing a stellar bulge, and a
Hernquist radial density profile \citep[][]{Hernquist}; the spin
parameter is set to $\lambda=0$. 
The mass ratio, $M_{\rm primary}/M_{\rm satellite}\sim 40$, is similar to
the estimated mass ratio of the Milky Way relative to
the Large Magellanic Cloud.
The main physical parameters of our simulated mergers are listed in Table~\ref{table:3}.

We consider prograde mergers, in which the satellite co-rotates with the
spin of the disk, as well as retrograde mergers, with a counter-rotating
satellite. We analyse two orbits: a low-inclination orbit with a
$10^{\circ}$ tilt with respect to the disk plane, and a high-inclination
orbit with a $60^{\circ}$ tilt. 

Initially, the particles of the small system (satellite galaxy) orbiting
around the (otherwise static) disk galaxy are all strongly
concentrated in space, and share essentially the same motion. 
The initial conditions (inclination, position, and velocity)
of the main system and the four impacting satellites, 
cosmologically motivated \citep[][]{Read, Villalobos}, 
are summarised in Table~\ref{table:4}. 

From the grid of simulations by \citet[][]{Read}, we chose 
four impactors, all of which having the mass of the Large Magellanic Cloud.
Larger masses would affect the stability of the stellar disk, 
and this is not consistent with a Milky Way-like galaxy.
Conversely, smaller masses would produce minor signatures 
in our local halo sample.

The four simulations are compiled using the public parallel Treecode
GADGET2 \citep{Springel} on the cluster matrix at the CASPUR 
({\it Consorzio Interuniversitario per le Applicazioni del Supercalcolo})
consortium, Rome. All systems were left to evolve for $4.63$~Gyr (about
$16$ dynamical timescales of the main halo). After this time, the four
satellites have completed their merging with the primary halo. The final
$(x,z)$ distribution of the inner satellite star particles and host disk, in
both the retrograde and prograde cases, as well as for the high and low
inclinations, is shown in Fig.~\ref{PRF:fig5}.

%________________________________________________________________
\begin{deluxetable}{lrlrrll}[hb!]
\tablewidth{0pt}
\tablecaption{\label{table:3} Physical properties of the Halos: Main system and orbiting Satellite.}
\tablehead{
\colhead{System}  & \colhead{$M_{DM}$}  & \colhead{$M_{*}$}  & 
\colhead{$N_{DM}$}  &  \colhead{$N_{*}$}  & \colhead{$r_0$}  &  \colhead{$r_{disk}$}
}
\startdata
Main       & $10^{12}$           &$5.7\times 10^{10}$&$10^{6}$ &$10^{6}$ & 4     & 20\\
Satellite  & $2.4\times 10^{10}$ &$2.4\times 10^{ 9}$&$1.1\times 10^{5}$ &$10^{5}$ & 0.709& ...\\
\enddata
\tablecomments{
  Column 1: Main galaxy/Impacting system. 
  Column 2: DM mass, in $M_{\sun}$. 
  Column 3: stellar mass for disk/bulge in the main/satellite, in $M_{\sun}$. 
  Column 4: DM particles.
  Column 5: stellar particles for disk/bulge in main/satellite.
  Column 6: disk scale radius for the main halo, in kpc; Hernquist scale radius for the satellite, in kpc.
  Column 7: disk truncation radius, in kpc.}
\end{deluxetable}

\begin{deluxetable*}{lrlrrrrrrr}
\tablewidth{0pt}
\tablecaption{\label{table:4} Initial conditions of the Main system and the four impacting Satellites.}
\tablehead{
\colhead{System}  & \colhead{Inclination}  & \colhead{Rotation} & \colhead{$x$}  & \colhead{$y$}  & \colhead{$z$}  & 
\colhead{$v_x$}  &  \colhead{$v_y$}  & \colhead{$v_z$}  &  \colhead{$v$}\\
& & & (kpc) & (kpc) & (kpc) & $\rm{(km~s^{-1})}$ & $\rm{(km~s^{-1})}$ & $\rm{(km~s^{-1})}$ & $\rm{(km~s^{-1})}$
}
\startdata
Main & $0^{\circ}$ & $-$ & 0.00& $0.00$ &  $0.00$ & $0.00$ &  $0.00$ &$0.00$ &   $0.00$ \\
Satellite 1  & $10^{\circ}$ & retrograde & $80.00$ & $0.27$ & $15.20$ &$6.30$ & $-62.50$ & $0.35$ & $62.82$\\
Satellite 2  & $10^{\circ}$ & prograde   & $80.00$ & $0.27$ & $15.20$ &$6.30$ &  $62.50$ & $0.35$ & $62.82$\\
Satellite 3  & $60^{\circ}$ & retrograde & $15.00$ & $0.12$ & $26.00$ &$1.20$ &  $80.10$ & $2.00$ & $80.13$\\
Satellite 4  & $60^{\circ}$ & prograde   & $15.00$ & $0.12$ & $26.00$ &$1.20$ & $-80.10$ & $2.00$ & $80.13$\\
\enddata
\tablecomments{Inclination and rotation of the orbit, position and velocity components, and total velocity.}
\end{deluxetable*}

\subsection{Dynamical friction and tidal stripping}\label{sec:4.2}

Any satellite can in principal be slowed by dynamical friction
exerted on it by disk and halo particles. It is known that an object,
such as a satellite, of mass $M$, moving through a homogeneous background
of individually much lighter particles with an isotropic velocity
distribution suffers a drag force \citep{Chandrasekar}:
$$F_d=-\frac{4\pi G^2M^2\rho_f(< v_s)\ln\Lambda}{v_s^2},$$ where $v_s$
is the speed of the satellite with respect to the mean velocity of the
field, $\rho_f(< v_s)$ is the total density of the field particles
with speeds less than $v_s$, and $\ln\Lambda$ is the Coulomb logarithm
\citep{BinneyTremaine}.

We expect that, the higher the $v_s$, the weaker is the dynamical
friction force. Retrograde satellites are expected to suffer weaker
dynamical friction with respect to prograde ones, since in the first
case the velocity of the satellite is opposite to that of the disk. 
As a consequence, prograde orbits decay faster. 
This effect is even more evident for low-inclination orbits.

Another important effect that occurs during mergers is the tidal
disruption of satellites. While tidal disruption is most important near
the centre of the main halo, where the gravitational potential is
changing more rapidly, dynamical friction is exerted both by the main
halo DM particles and by the disk star particles.           

   \begin{figure*}
   \centering
   \includegraphics[width=\linewidth]{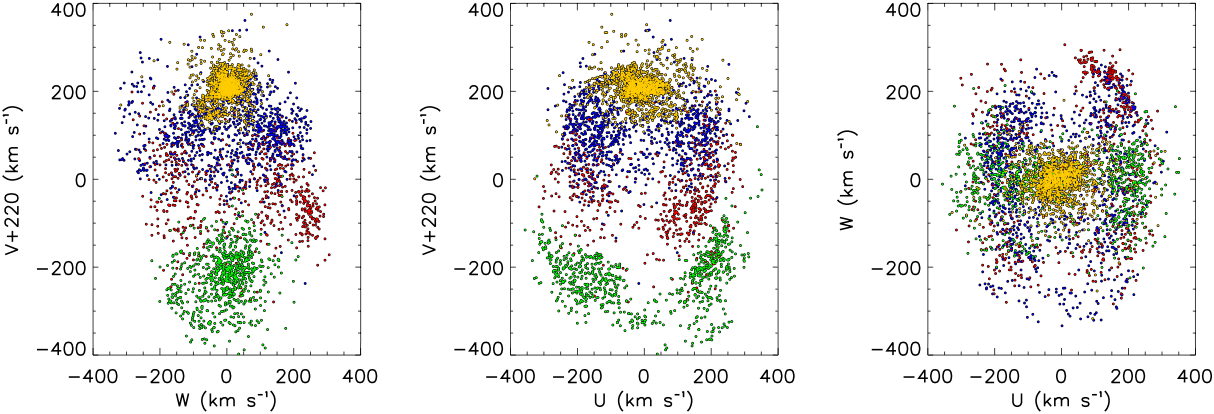}
   \caption{Kinematical (velocity space) distribution of the 
	accreted component of the simulated Milky Way inner halo, 
     i.e., $3902$ particles in a spherical volume of radius $3$~kpc centered on the ``Sun" with $|z|>1$~kpc. 
     Different colours indicate particles associated with different satellites: 
     $60^{\circ}$ retrograde/prograde (red/blue),
     $10^{\circ}$ retrograde/prograde (green/yellow) colliding satellites.}
              \label{PRF:fig6}
    \end{figure*}
%________________________________________________________________

\subsection{Debris in the local halo}\label{sec:4.3}

We analyse the observational signature left by the satellite stars after selecting particles in a sphere of $3$~kpc
radius centered at the Sun ($x=8$~kpc from the Galaxy center), and with 
$|z|>1$~kpc. This last constraint is introduced 
so that the simulated and observed samples can be compared within 
a similar volume of the inner halo.

Figure~\ref{PRF:fig6} shows the kinematic distribution (velocity
projections) of our simulated inner halo. 
The different colours indicate the association of 
the 3902 debris stars with different progenitors: the low/high-inclination
retrograde satellites (761/616 green/red dots), and the 
high/low-inclination prograde satellites (966/1559 blue/yellow dots).\\

\begin{figure*}
   \centering
   \includegraphics[width=\linewidth]{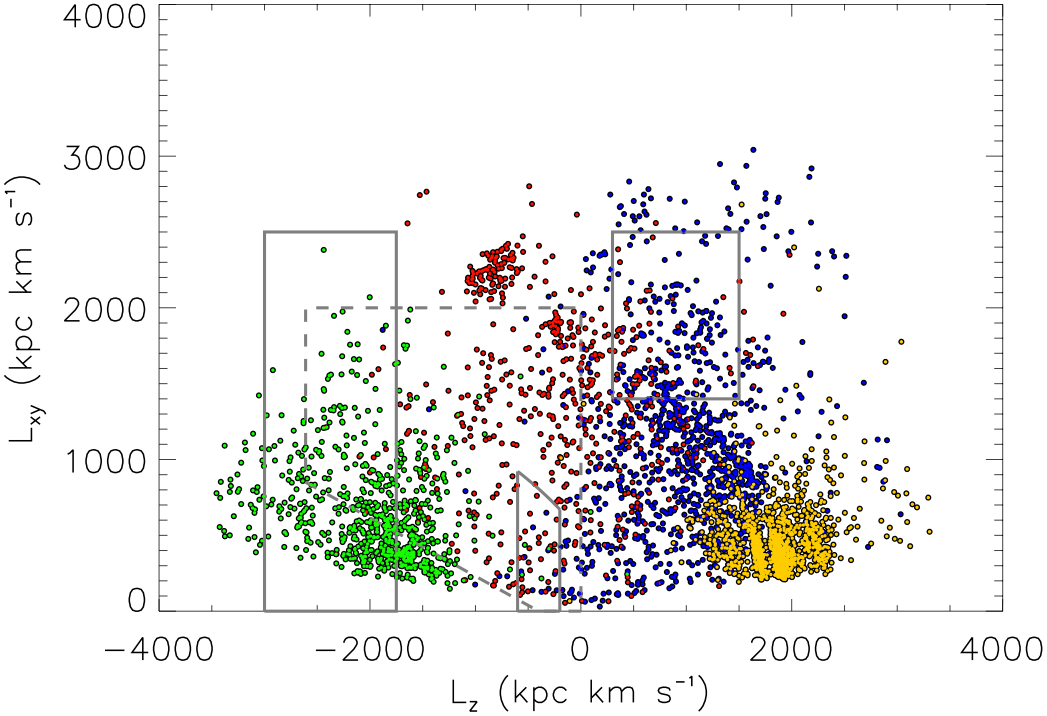}
      \caption{Angular momentum distribution of the simulated Milky Way halo within $3$~kpc of the ``Sun".
        	As in Fig.~\ref{PRF:fig6}, shown are the $3902$ particles accreted from four dwarf galaxies:
        	$60^{\circ}$ retrograde/prograde (red/blue), $10^{\circ}$ retrograde/prograde (green/yellow) satellites
        	after interaction with the simulated Milky Way.
		All of the marked regions 
		have the same meaning as in Fig.~\ref{PRF:fig4}. 
              }
         \label{PRF:fig7}
   \end{figure*}

The angular momentum distribution of the
satellite debris is shown in Fig.~\ref{PRF:fig7}.  
Despite the chaotic build up of the parent halo, 
it appears that objects from accreted satellites remain confined in limited portions of 
the $(L_z, L_{xy})$ plane.  

The satellite on a low-inclination prograde orbit (yellow particles), 
which suffers more from dynamical friction, quickly loses its
orbital energy and proceeds to the inner regions of the main halo
\citep{Byrd, Murante}. 
For this reason less particles are left in the outer-halo, 
see top-right panel of Fig.~\ref{PRF:fig5}. 

It is worth noticing that the high inclination prograde satellite 
suffers the effect of dynamical friction as well, as a result of its co-rotation with the disk. 
This effect acts in producing a consistent mass of debris in the solar region having 
$L_{xy}$ ranging between $500$~$\rm{kpc~km~s^{-1}}$ and $1500$~$\rm{kpc~km~s^{-1}}$ 
(blue points in Fig.~\ref{PRF:fig7}). 

On the other hand, retrograde satellites experience weaker dynamical
friction and leave more particles in the outer-halo region, since
their orbits have a longer decay time and longer periods. Thus, tidal
stripping \citep[see e.g.,][]{Colpi} can act longer and more efficiently
when the satellite is still orbiting at high velocity, and we see that a
better populated high-velocity tail results (compare Fig.~\ref{PRF:fig7} to Fig.~\ref{PRF:fig10}).

The impact of dynamical friction on the two configurations considered
for the retrograde satellites indicates that the high-inclination case
is the one less affected by this force, which again results in efficient
stripping when the satellite has high orbital velocity.  
Therefore, such stripping takes place over a large spatial region, 
and for the conservation of 6D phase-space density, 
by virtue of Liouville's Theorem,  
we expect a small variance in velocity space and in the plane of angular momenta. 
This is indeed observed for the red particles with respect to the green ones 
in Figs.~\ref{PRF:fig6} and~\ref{PRF:fig7}.

Finally, the effect of both gravitational feedback and dynamical friction 
on the satellites, which lead to loss of stars at different passages with different energies, 
is clearly evident for the case of low inclination prograde orbit in Fig.~\ref{PRF:fig7} at around 
$L_{xy}=400~\rm{kpc~km~s^{-1}}$ and $L_{z}=1750~\rm{kpc~km~s^{-1}}$.

%________________________________________________________________

\section{``Observed" simulations}\label{sec:5}

Here we investigate the effects of observational errors on our
simulated data, and show how more accurate kinematic data
to be provided by future surveys can improve detection and characterisation 
of halo streams. 
Moreover, we compare {\it actual} observations with the
distributions of debris resulting from the four 
simulated satellites presented in the previous section, 
and discuss the orbital properties of the parent dwarf galaxies,
possibly responsible for accreting on the Milky Way halo.

%________________________________________________________________

\subsection{Observational errors}\label{sec:5.1}

   \begin{figure*}
   \centering
   \includegraphics[angle=0, width=1.\linewidth]{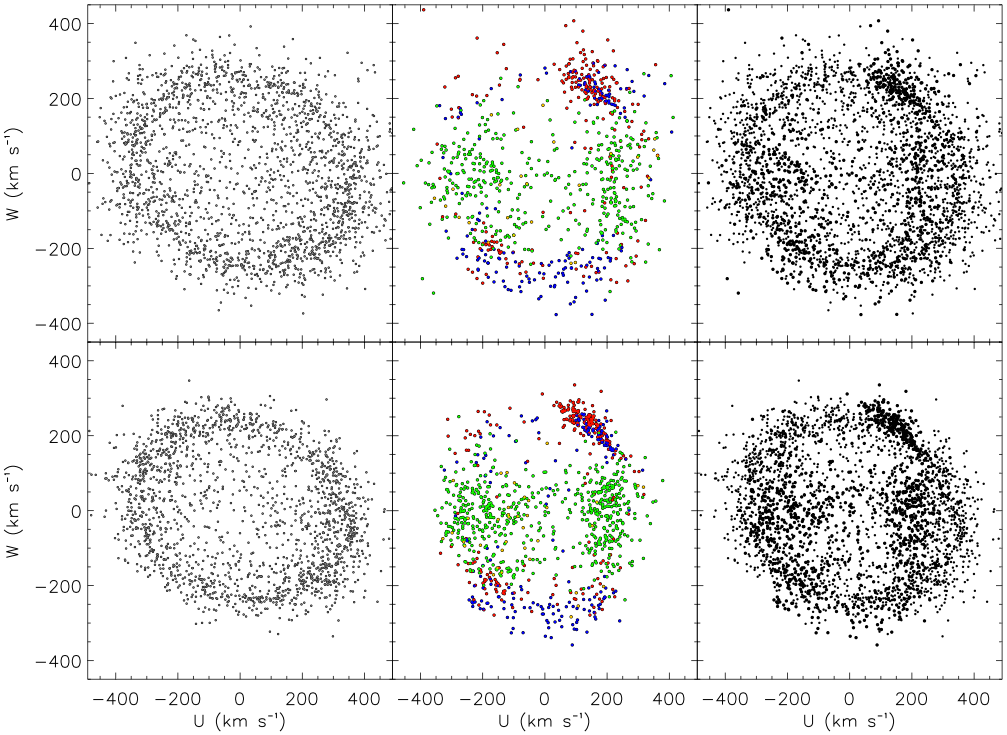}
   \caption{
The ($U$, $W$) velocity distribution of the $10\%$ high velocity tail of the simulated sample is shown in the right panels. The sample is limited to a spherical volume of radius 3 kpc located on the plane of the simulated Milky Way 8 kpc away from its center and with $|z|>1$~kpc. Of the 2874 particles in this sample, some are 
remnants of the four satellites accreted after 5 Gyr 
(middle panels), the remaining constitute the ``background" of smooth inner halo stars (left panels).
Different colours represent different progenitors as in Fig.~\ref{PRF:fig5}. 
Finally, the top panels were generated via convolution with current, i.e. SGKS, ground-based errors, while the bottom figures depict the results after convolution with the expected Gaia-like errors.
}
	\label{PRF:fig8}  
    \end{figure*}

We perturb the original simulations by convolving the ``true"
data with two cases of error distributions. 
First, we adopt the accuracy of our SGKS catalogue as representative of the quality of 
current wide-field surveys.
Then, we assume the mean accuracy expected from the forthcoming Gaia
catalogue combined with complementary deep spectroscopic data from on-going and future surveys such as GES 
\citep[][]{Gilmore}.

The true positions and velocities of each particle are first transformed
into their astronomical observables ($\alpha$, $\delta$, $m - M$, or $\pi$ and
$\mu_\alpha$, $\mu_\delta$, $V_r$); then the expected observational errors
are added to distance modulus (or directly to parallax, 
in the case of the Gaia-like simulation), 
radial velocity, and proper motion, according to Table~\ref{table:5}. 

The precision in distance is taken to $\sigma_{m - M} = 0.4$~mag (i.e., 
$\sigma_d/d \simeq 20\%$) for the photometric distances estimated from the SGKS 
catalogue, and to $\sigma_\pi=20~\rm{\mu as}$ for the
final precision on trigonometric parallaxes measured by Gaia. 
In proper motion, the precision is assumed to be $2~\rm{mas~yr^{-1}}$ for ground-based
observations, and $20~\rm{\mu as~yr^{-1}}$ for Gaia. The precision in
the radial velocity is taken to be $10~\rm{km~s^{-1}}$ for the SDSS measurements, 
and $1~\rm{km~s^{-1}}$ for the GES spectroscopic survey. These quantities
are finally transformed back to observed positions vectors and space velocities.

\begin{deluxetable*}{ccccc}
\tablewidth{0pt}
\tablecaption{\label{table:5} Estimated/Expected errors for the SGKS and Gaia catalogues.}
\tablehead{
\colhead{Catalogue} & \colhead{distance}  & \colhead{proper motion}  & \colhead{radial velocity} 
%              &          & $\rm{(\mu as~yr^{-1})}$ & $\rm{(km~s^{-1})}$ 
}
\startdata
   SGKS       & $\sigma_{m - M}=0.4$ mag      & 2000 & 10\\ 
   Gaia/GES   & $\sigma_\pi=20~\rm{\mu as}$   &   20 &  1\\
\enddata
\tablecomments{
Estimated errors (precision) in parallax ($\sigma_\pi$, in $\rm{\mu as}$), distance modulus ($\sigma_{m - M}$, in mag), 
proper motion ($\sigma_\mu$, in $\rm{\mu as~yr^{-1}}$), and
radial velocity ($\sigma_{V_r}$, in $\rm{km~s^{-1}}$) 
for the SGKS and Gaia catalogues.}
\end{deluxetable*}

\begin{deluxetable*}{lcrrrrr}
\tablewidth{0pt}
\tablecaption{\label{table:6} The composition of the 10\% High Velocity Tail of the simulated Milky Way Halo.}
\tablehead{
\colhead{Catalogue} & \colhead{Halo$+$Debris} & \colhead{Debris} & \colhead{Satellite 1} & \colhead{Satellite 2} & \colhead{Satellite 3} & \colhead{Satellite 4}
}
\startdata
   ``True" Simulation   & 2874   &     $1103~(0.38\%)$  &       262   &      201   &      601    &      39 \\ 
   ``Observed" SGKS     & 2874   &     $ 835~(0.29\%)$  &       233   &      170   &      406    &      26 \\  
   ``Observed" Gaia/GES & 2874   &     $1061~(0.37\%)$  &       263   &      191   &      570    &      37 \\
\enddata
%\tablecomments{The simulation}
\end{deluxetable*}

%________________________________________________________________

\subsection{The inner halo model}\label{sec:5.2}

We explore a simulated inner halo based on 
the superposition of the four 
simulations of minor mergers  
and a smooth local component with the same kinematic properties of the observed sample (Table~\ref{table:1}).

Consistently with the findings of \citet[][]{Helmi} and \citet[][]{Kepley}, 
we assumed a debris total fraction of $10\%$ within $3$~kpc from the Sun.

In the ``true" (simulated) catalogue, of the $28\,738$ particles with $|z|>1$~kpc, 
$24\,836$ are part of the local mock halo, while the remaining $3902$ are debris from the satellites 
shown in Figs.~\ref{PRF:fig6} and~\ref{PRF:fig7}.

In the following discussion 
we focus on the 
particles of the accreted component that belong to the $10\%$ 
high velocity tail. 
Table~\ref{table:6} reports the number of particles belonging to each satellite 
for the pure simulation and the other two cases accounting for observational errors.

Figure~\ref{PRF:fig8} shows the region of the $(U, W)$ plane 
occupied by the $10\%$ high velocity tail of the resulting simulated Milky Way inner halo (right panels), 
as superposition the accreted component (middle panels) and the smooth spheroid (left panels). 
The synthetic ``observed" catalogue shown in 
the top panels 
represent the current picture, 
according to the SGKS error model. 
The bottom panels show the distribution of the high velocity particles as promised by Gaia. 

The upper panels indicate that distinguishing in velocity space the satellites
that gave rise to each of the different moving groups with the extant data is a
non-trivial task. 
On the other hand, as the inspection of the lower panels reveals, 
much of the 
substructures shown in the middle panels 
becomes visible again thanks to the superior precision that Gaia will achieve.

%________________________________________________________________

\subsection{Substructures in the correlation function}\label{sec:5.3}

In order to quantify the amount of kinematic substructures present  
among the 2874 fastest-moving particles, 
we
compute the cumulative velocity correlation function described in
Sect.~\ref{sec:3.2}.
The analysis is performed over three synthetic catalogues: 
the ``true" simulation, and two lists derived from the true values after perturbing them with 
either SGKS-like errors 
or the errors expected for the Gaia/GES surveys.

 \begin{figure}[ht!]                                              
   \centering
   \includegraphics[width=\linewidth]{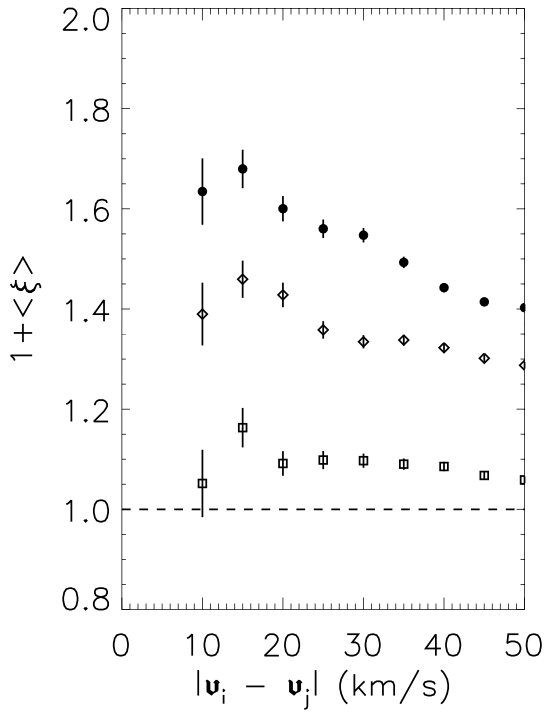}
   \caption{
	The cumulative velocity 
	correlation functions for the 2874 halo particles 
	shown in Fig.~\ref{PRF:fig8}.
	The filled dots trace the correlation function of the pure simulation, 
	while squares and diamonds depict the correlation after convolution with 
	current (i.e., SGKS) or Gaia/GES-like errors, respectively.
	Error bars are derived from Poisson's statistics of the counts.
	}
         \label{PRF:fig9}
   \end{figure}

Figure~\ref{PRF:fig9} shows,
using bins of width $\rm{5~km~s^{-1}}$ up to kinematical separations of $\rm{50~km~s^{-1}}$, 
the results for the two-point correlation function $\xi({\bf v})$ for the 
``true" case (dots), the SGKS-like (squares) and Gaia/GES-like (diamonds) catalogues.
The clear signal in the first bins, peaking at $\Delta v \sim \rm{15~km~s^{-1}}$, evidences an excess of particles moving with similar velocities with respect to what expected from a fully random sample. 
In the case of the pure simulation, 
the two-point velocity correlation function attains a maximum signature of %$\sim 65\%$, with a SNR$>10$ 
$\langle \xi \rangle = 0.68 \pm 0.04$.
For the SGKS-like catalogue the maximum signal 
has an $\sim 80$\% drop 
to an ``observed" value of $\langle \xi \rangle = 0.16 \pm 0.04$.

The recovery of the intrinsic correlation signal is truly remarkable when looking at the 
correlation function 
for the Gaia-like case: 
in fact, we measure $\langle \xi \rangle = 0.46 \pm 0.04$, corresponding to 
$68\%$ of the original signal. 
Figure~\ref{PRF:fig8} provides a nice ``visual" confirmation of the 
recovery in substructure visibility.

%________________________________________________________________
%\newpage

\section{On the nature of the high velocity debris}\label{sec:6}

\begin{figure*}
   \centering
	\includegraphics[width=\linewidth]{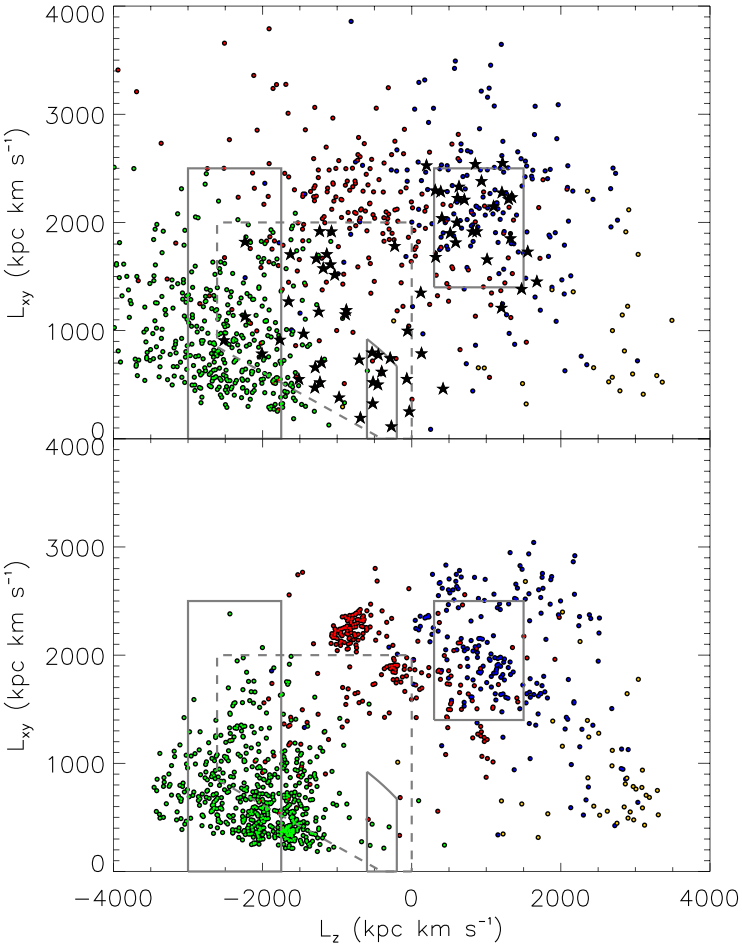}
      \caption{
        The $10\%$ high velocity tail component belonging solely to the 4 satellites; 
        convolved with current ground-based errors (top, 835 particles) 
		and with the expected Gaia errors (bottom, 1061 particles).
		Black star symbols are the $67$ fast moving debris stars uncovered from the analysis of the 
		SGKS sample.
		Solid and dashed contours have the same meaning as in Figs.~\ref{PRF:fig4} and~\ref{PRF:fig7} .
              }
         \label{PRF:fig10}
   \end{figure*}

As the space of adiabatic invariants is important to gain more insight
into the properties of the kinematic substructures detected (Sect.~\ref{sec:3.4}), 
we compare the $(L_z,L_{xy})$ distributions of the observed groups with the results of the simulations, 
taking into account the effect of the observational errors.
This is shown in Fig.~\ref{PRF:fig10}, 
where the top panel corresponds to the SGKS-error simulation, while the bottom panel 
reproduces what will hopefully be seen with the final Gaia catalogue.

The black star symbols in the upper panel of Fig.~\ref{PRF:fig10} 
represent the $67$ high velocity objects we found from our statistical analysis in the same volume 
and shown in Fig.~\ref{PRF:fig4} as colored stars. 
With current data, different satellites mix over some regions so that a
discrete classification is not always straightforward. 
The bottom panel of Fig.~\ref{PRF:fig10} clearly shows that this situation is highly improved with Gaia-like data.\\

We see that our Groups~1 and 2, corresponding to the stream of \cite{Helmi}, 
are consistently associated with the high inclination prograde satellite (blue dots).
Because of dynamical friction (cfr. Sect.~\ref{sec:4.3}), this
satellite includes a low $L_{xy}$ component shown in the full sample (Fig.~\ref{PRF:fig7}) 
that is not part of the high velocity tail (Fig.~\ref{PRF:fig10}). 
Thus, these simulated ``observations" suggest 
the possible presence in the 
\citet{Helmi} stream of debris with lower $L_{xy}$ yet to be discovered.

Of particular interest is the case of the retrograde kinematic groups. 
In fact, neither the high-inclination simulated satellite nor the one at low-inclination appear 
to fairly match the observed Groups 3, 4, and 5, i.e. the black stars with $L_z \la 0$ in Fig.~\ref{PRF:fig10}. 

Actually, these three groups appear to occupy an {\it intermediate} region 
between the debris of the two simulated retrograde satellites.
Furthermore, in Sect.~\ref{sec:3.4} we remark that 
the observed Groups~3, 4, and 5 do not well match the streams detected by 
\citet{Dinescu} and \citet{Kepley}.
For this reason, we suggest that these three groups may represent 
the debris of an unique progenitor accreted along an initial retrograde orbit having 
an intermediate inclination in the range comprised between $10^{\circ}$ and $60^{\circ}$. 
In alternative, these groups could belong up to 
three different impacting satellites on retrograde orbits 
with inclinations in that same range.

The results presented in this section show that the methodology proposed is 
certainly capable of detecting fossil signatures as kinematic substructures among 
high-velocity stars.
From the data at our disposal, there is clear indication that 
more debris are found from dwarf 
galaxies on high-inclination prograde and retrograde orbits,  
as well as on low-inclination retrograde ones. 
We have not identified any debris coming from low-inclination prograde
satellites and this might 
be a limitation intrinsic to the methodology of 
looking at structures in the space motions of 
very high velocity stars. 
Future work will have to investigate these issues.
 
%________________________________________________________________

\section{Conclusions}\label{sec:7}

We have explored the Solar neighborhood of the Milky Way through the
use of spectro-photometric data from the Sloan Digital Sky Survey and
high-quality proper motions derived from multi-epoch positions extracted
from the Guide Star Catalogue II database.
A sample with accurate distances, space velocities, and metallicities is selected 
as a tracer of the inner-halo population resulting in 
$2417$ subdwarfs with $\rm{[Fe/H]}<-1.5$ and $\rm{|z|>1}$~kpc
within $3$~kpc of the Sun.
This set is then analysed to identify and characterise kinematic streams possibly 
arising from merging events.

We have found statistical evidence of substructures in the space motions
of the $10\%$ fastest stars, confirming the existence of $5$ moving groups.

In angular momenta space, the two prograde groups we have identified 
(Groups 1 and 2 in Table~\ref{table:2}) 
appear confined in the region encompassing the stream first identified by \cite{Helmi} 
among red giants and RR Lyrae within $1$~kpc of the Sun. 
Our analysis found $25$ additional subdwarf members belonging to that same stream: % Helmi stream: 
$4$ are in common with those found by \cite{Klement}, while the other $21$  
are newly discovered members.

Of the remaining groups, the most counter-rotating one (Group~4) partially overlaps with 
the region of ``retrograde outliers" found by \citet{Kepley},
while a dozen stars belonging to Group~4 and Group~5 fall in the region 
of the mildly retrograde stream detected by \cite{Dinescu}. 

Comparison to our high resolution N-body 
simulations confirms 
that the two groups associated with the Helmi stream are 
likely fossil remnants of a dwarf galaxy which co-rotates with the disk of the 
Galaxy and moves on a high-inclination orbit. 

As for the three retrograde groups (3, 4 and 5 in Table~\ref{table:2}), they may be 
debris of an unique progenitor accreted along an initial retrograde
orbit having an intermediate inclination  in the range $10^{\circ} \div 60^{\circ}$. 
However, we cannot exclude that these groups belong to different impacting satellites on retrograde
orbits with inclinations within that same range.
A more detailed analysis of the chemical abundances of the three detected groups,  
as well as more quantitative comparisons to extended simulations, are 
necessary to resolve this issue.

In any event, the fastest objects appear with positive and negative $L_z$ values 
(i.e. prograde and retrograde motions, respectively)
in the angular momentum  $(L_z, L_{xy})$ regions 
for high-inclination orbits 
($L_{xy} \gtrsim 1500~\rm{kpc~km~s^{-1}}$).
On the other hand, for low-inclination  
both observations and simulations 
show that the fastest objects 
appear only on {\it retrograde} orbits
(e.g., Fig.~\ref{PRF:fig10}, top panel).
This asymmetric distribution is suggestive of the role played by
dynamical friction during accretion. 

In anticipation of the much improved data expected over the coming years, in particular the Gaia catalogue 
and the new ground-based spectroscopic surveys, we also investigated the impact of 
observational errors in our dynamical simulations.
The analysis indicates that 
(see the relevant panels of Figs.~\ref{PRF:fig8} and~\ref{PRF:fig10})
Gaia will greatly influence these studies: 
for, velocity and angular momentum distribution will be almost completely dominated by the physics we are trying to recover, 
i.e., the dynamical history of the merging events.

At that point, 
full grids (in, e.g., inclination and amount of rotation) 
of prograde and retrograde high resolution satellite simulations 
will be required to precisely characterise the debris detected.
Then, we will be able to number the merging events 
for direct comparison with the predictions of the ($\Lambda$)CDM theory and its associated merging paradigm. 

In conclusion, the results shown might lead us to claim that the inner halo might have ``seen" only 
two events; 
however the large uncertainties in the extant data, 
mostly observational, do not exclude the possibility that the events might be as many as four, and perhaps more given 
the intrinsic difficulty of our technique to deal with low inclination prograde mergers.

%________________________________________________________________

\begin{acknowledgements}

We are grateful to the referee for his/her comments that helped us improve the original manuscript.
This work has been partially funded by ASI, under contract to INAF I/058/10/0 
``Gaia Mission - The Italian Participation to DPAC", and by MIUR, through PRIN 2012 grant No 1.05.01.97.02 
``Chemical and dynamical evolution of our Galaxy and of the galaxies of the Local Group". 
M.G.L. acknowledges support from the Chinese Academy of Sciences through 2015 
CAS President's International Fellowship Initiative (PIFI) for Visiting Scientists.

\end{acknowledgements}

%________________________________________________________________


\begin{thebibliography}{}

\bibitem[Abadi et~al. (2006)]{Abadi} Abadi, M. G., Navarro, J. F., \& Steinmetz, M. 2006, MNRAS, 365, 747
\bibitem[Abazajian et~al. (2009)]{SDSS7} Abazajian, K. N., Adelman-McCarthy, J. K., Ag\"ueros, M. A., et al. 2010, ApJS, 182, 543
\bibitem[Allende Prieto et~al. (2008)]{SSPP3} Allende Prieto, C., Sivarani, T., Beers, T. C., et al. 2008, AJ, 136, 2070

\bibitem[Byrd et~al. (1986)]{Byrd} Byrd, G. G., Saarinen, S. \& Valtonen, M. J. 1986, MNRAS, 220, 619

\bibitem[Bullock \& Johnston (2005)]{BullockJohnston} Bullock, J. S., \& Johnston K. V. 2005, ApJ, 635, 931 
\bibitem[Binney \& Merrifield (1998)]{BinneyMerrifield} Binney, J., \& Merrifield, M. 1998, Galactic Astronomy (Princeton Univ. Press, Princeton)
\bibitem[Binney \& Tremaine (1987)]{BinneyTremaine} Binney, J., \& Tremaine, S. 1987, Galactic Dynamics (Princeton Univ. Press, Princeton)
\bibitem[Bond et~al. (2010)]{Bond} Bond, N. A., Ivezi\'{c}, \v{Z}., Sesar, B., et al. 2010, ApJ, 716, 1
\bibitem[Chandrasekar (1943)]{Chandrasekar} Chandrasekar, S. 1943, ApJ, 97, 255
\bibitem[Chiba \& Beers (2000)]{ChibaBeers} Chiba, M., \& Beers, T. C. 2000, AJ, 119, 2843
\bibitem[Colpi et~al. (1999)]{Colpi} Colpi, M., Mayer, L., \& Governato, F. 1999, ApJ, 525, 720
\bibitem[Dehnen \& Binney (1998)]{DehnenBinney} Dehnen, W., \& Binney, J. J. 1998, MNRAS, 298, 387
\bibitem[De Lucia (2012)]{DeLucia} De Lucia, G. 2012, Astron. Nachr., 333, 460 
\bibitem[Dinescu (2002)]{Dinescu} Dinescu, D. I. 2002, in: F. van Leeuwen, J. Hughes, \& G. Piotto (eds.), $\Omega$ Centauri: A Unique Window into Astrophysics, (San Francisco: ASP), Vol. 265, p. 365
\bibitem[Duffau et~al. (2014)]{Duffau} Duffau, S., Vivas, K. A., Zinn, R., Mendez, R. A., \& Ruiz, M. T. 2014, A\&A, 566, 118
\bibitem[Eggen (1977)]{Eggen} Eggen, O. J. 1977, ApJ, 215, 812
\bibitem[Eggen (1971)]{Eggen71} Eggen, O. J. 1971, PASP, 83, 285
\bibitem[Freeman \& Bland-Hawthorn (2002)]{FreemanBland-Hawthorn} Freeman, K., \& Bland-Hawthorn, J. 2002, ARA\&A, 40, 487
\bibitem[Gilmore et~al. (2012)]{Gilmore} Gilmore, G., Randich, S., Asplund, M., et al. 2012, The Messenger, 147, 25
\bibitem[Girardi et~al. (2004)]{Girardi} Girardi, L., Grebel, E. K., Odenkirken, M., \& Chiosi, C. 2004, A\&A, 422, 205
\bibitem[Gomez et~al. (2013)]{Gomez} Gomez, F. A., Helmi, A., Cooper, A. P., et al. 2013, MNRAS, 436, 3602
\bibitem[Gould (2003)]{Gould} Gould, A. 2003, ApJ, 592, L63
\bibitem[Harding et~al. (2001)]{Harding} Harding, P., Morrison, H. L., Olszewski, E. W., et al. 2001, AJ, 122, 1397
\bibitem[Hastie et~al. (2001)]{Hastie} Hastie, T., Tibshirani, R., \& Friedman, J. 2001, The Elements of Statistical Learning (Springer)
\bibitem[Helmi et~al. (2011)]{Helmi11} Helmi, A., Cooper, A. P., White, S. D. M., et al. 2011, ApJ, 733, L7
\bibitem[Helmi (2008)]{HelmiRev} Helmi, A. 2008, A\&AR, 15, 145
\bibitem[Helmi \& de Zeeuw (2000)]{HelmideZeeuw} Helmi, A., \& de Zeeuw, P. T. 2000, MNRAS, 319, 657
\bibitem[Helmi et~al. (1999)]{Helmi} Helmi, A., White, S. D. M., de Zeeuw, P. T., \& Zhao, H. S. 1999, Nature, 402, 53
\bibitem[Helmi \& White (1999)]{HelmiWhite} Helmi, A., \& White, S. D. M. 1999, MNRAS, 307, 495 
\bibitem[Hernquist (1993)]{Hernquist} Hernquist, L. 1993, ApJS, 86, 389 
\bibitem[Ibata et~al. (2003)]{Ibata03} Ibata, R. A., Irwin, M. J., Lewis, G. F., Ferguson, A. M. N., \& Tanvir, N. 2003, MNRAS, 340, 21
\bibitem[Ibata et~al. (1994)]{Ibata94} Ibata, R. A., Gilmore, G., \& Irwin, M. J. 1994, Nature, 370, 194
\bibitem[Ivezi\'{c} et~al. (2008)]{Ivezic} Ivezi\'{c}, \v{Z}., Sesar, B., Juri\'{c}, M., et al. 2008, ApJ, 684, 287
\bibitem[Johnston (1998)]{Johnston98} Johnston, K. V. 1998, ApJ, 495, 297
\bibitem[Kaufmann \& Rousseeuw (1990)]{Kaufmann} Kaufmann, L.,  \& Rousseeuw, P. J. 1990, Finding Groups in Data (John Wiley \& Sons, Inc.)
\bibitem[Kepley et~al. (2007)]{Kepley} Kepley, A. A., Morrison, H. L., Helmi, A., et al. 2007, AJ, 134, 1579
\bibitem[Klement (2010)]{KlementRev} Klement, R. J. 2010, A\&AR, 18, 567
\bibitem[Klement et~al. (2009)]{Klement} Klement, R., Rix, H. -W., Flynn, C., et al. 2009, ApJ, 698, 865 
\bibitem[Lasker et~al. (2008)]{Lasker} Lasker, B. M., Lattanzi, M. G., McLean, B. J., et al. 2008, AJ, 139, 735
\bibitem[Lee et~al. (2008a)]{LeeSSPP1} Lee, Y. S., Beers, T. C., Sivarani, T., et al. 2008a, AJ, 136, 2022
\bibitem[Lee et~al. (2008b)]{LeeSSPP2} Lee, Y. S., Beers, T. C., Sivarani, T., et al. 2008b, AJ, 136, 2050
\bibitem[Majewski et~al. (2012)]{Majewski2012} Majewski, S. R., Nidever, D. L., Smith, V. V., et al. 2012, ApJ, 747, L37
\bibitem[Majewski et~al. (1996)]{Majewski} Majewski, S. R., Munn, J. E., \& Hawley, S. L. 1996, ApJ, 459, 73
\bibitem[Meza et~al. (2005)]{Meza} Meza, A., Navarro, J. F., Abadi, M. G., \& Steinmetz, M. 2005, MNRAS, 359, 93
\bibitem[Moore et~al. (2006)]{Moore} Moore, B., Diemand, J., Madau, P., Zemp, M., \& Stadel, J. 2006, MNRAS, 368, 563
\bibitem[Morrison et~al. (2009)]{Morrison} Morrison, H. L., Helmi, A., Sun, J., et al. 2009, ApJ, 694, 130
\bibitem[Munn et~al. (2008)]{Munn} Munn, J. A., Monet, D. G., Levine, S. E., et al. 2008, AJ, 136, 895
%\bibitem[Munn et~al. (2004)]{Munn} Munn, J. A., Monet, D. G., Levine, S. E., et al. 2004, AJ, 127, 3034
\bibitem[Murante et~al. (2010)]{Murante} Murante, G., Poglio, E., Curir, A., \& Villalobos, A. 2010, ApJ, 716, L115
\bibitem[Navarrete et~al. (2015)]{Navarrete} Navarrete, C., Chaname, J., Meza, A., et al. 2015, to appear in ApJ
\bibitem[Navarro, Frenk \& White (1997)]{Navarro} Navarro, J. F., Frenk, C. S., \& White, S. D. M. 1997, ApJ, 490, 493
\bibitem[Pasetto et~al. (2012)]{Pasetto} Pasetto, S., Grebel, E. K., Zwitter, T., et al. 2012, A\&A, 547, A70
\bibitem[Perryman et~al. (2001)]{Perryman} Perryman, M. A. C., de Boer, K. S., Gilmore, G., et al. 2001, A\&A, 369, 339
\bibitem[Read et~al. (2008)]{Read} Read, J. I., Lake, G., Agertz, O., \& Debattista, V. P. 2008, MNRAS, 389, 1041
\bibitem[Re~Fiorentin et~al. (2007)]{ReFiorentin07} Re Fiorentin, P., Bailer-Jones, C. A. L., Lee, Y. S., et al. 2007, A\&A, 467, 1373
\bibitem[Re~Fiorentin et~al. (2005)]{ReFiorentin05} Re Fiorentin, P., Helmi, A., Lattanzi, M. G., \& Spagna, A. 2005, A\&A, 439, 551
\bibitem[Sales et~al. (2007)]{Sales} Sales, L. V., Navarro, J. F., Abadi, M. G., \& Steinmetz, M. 2007, MNRAS, 379, 1464
\bibitem[Schlaufman et~al. (2009)]{Schlaufman09} Schlaufman, K.C., Rockosi, C. M., Allende Prieto, C., Beers, T. C., Bizyaev, D., et al. 2009, ApJ, 703, 2177
\bibitem[Schlaufman et~al. (2011)]{Schlaufman11} Schlaufman, K.C., Rockosi, C. M., Lee, Y.S., Beers, T.C., \& Allende Prieto, C. 2011, ApJ, 734, 49
\bibitem[Schlafly et~al. (2014)]{Schlafly2014} Schlafly, E. F., Green, G., Finkbeiner, D. P., et al. 2014, ApJ, 789, 15
\bibitem[Schlafly \& Finkbeiner (2011)]{Schlafly} Schlafly, E. F., \& Finkbeiner, D. P. 2011, ApJ, 737, 103
\bibitem[Schlegel et~al. (1998)]{Schlegel} Schlegel, D. J., Finkbeiner, D. P., \& Davis, M. 1998, ApJ, 500, 525 
\bibitem[Searle \& Zinn (1978)]{SZ} Searle, L., \& Zinn, R. 1978, ApJ, 225, 357
\bibitem[Smith et~al. (2009)]{Smith2009} Smith, M. C., Evans, N. W., Belokurov, V., et al. 2009, MNRAS, 399, 1223
\bibitem[Spagna et~al. (2010a)]{SpagnaL} Spagna, A., Lattanzi, M. G., Re~Fiorentin, P., \& Smart, R. L. 2010a, A\&A, 510, L4
\bibitem[Spagna et~al. (2010b)]{Spagna10} Spagna, A., Bucciarelli, B., Lattanzi, M. G., Re~Fiorentin, P., \& Smart, R. L. 2010b, MemSAIt, 14, 67
\bibitem[Spagna et~al. (2004)]{Spagna04} Spagna, A., Carollo, D., Lattanzi, M. G., \& Bucciarelli, B. 2004, A\&A, 428, 451
\bibitem[Springel (2005)]{Springel} Springel, V. 2005, MNRAS, 364, 1105
\bibitem[Starkenburg et~al. (2009)]{Starkenburg} Starkenburg, E., Helmi, A., Morrison, H. L., et al. 2009, ApJ, 698, 567
\bibitem[Trumpler \& Weaver (1953)]{Trumpler} Trumpler R. J., \& Weaver H. F. 1953, Statistical Astronomy (Dover Pubblications, Inc., New York)
\bibitem[Turon et~al. (2005)]{Turon} Turon, C., O'Flaherty, K. S., \& Perryman, M. A. C. eds. 2005, The Three-dimensional universe with Gaia, ESA Spec. Publ., 576
\bibitem[Villalobos \& Helmi (2008)]{Villalobos} Villalobos, A., \& Helmi, A. 2008, MNRAS, 391, 1806
\bibitem[Yanny et~al. (2009)]{Yanny} Yanny, B., Rockosi, C., Newberg, H. J., et al. 2009, AJ, 137, 4377
\bibitem[Zhao et~al. (2012)]{LAMOST} Zhao, G., Zhao, Y.-H., Chu, Y.-Q., et al. 2012, RAA, 12, 723


\end{thebibliography}
\end{document}